\documentclass[journal=jctcce,manuscript=article]{achemso}
\setkeys{acs}{articletitle=true}

\usepackage[utf8]{inputenc}
\usepackage{amssymb}
\usepackage{amstext}
\usepackage{amsmath}
\usepackage{graphicx}
\usepackage{color}
\usepackage{epsfig}
\usepackage[normalem]{ulem}
\usepackage{braket}
\usepackage{dsfont}
\usepackage{mathtools}
\usepackage{physics}
\usepackage{graphicx}
\usepackage{threeparttable}

\title{Using Non-covalent Interactions to Test Precision of Projector-Augmented Wave Data Sets}
\author{Sirous Yourdkhani}
\email{yourdkhani.sirous@karlov.mff.cuni.cz} 
\affiliation{Department of Chemical Physics and Optics, Faculty of Mathematics and Physics, Charles University, CZ-12116 Prague 2, Czech Republic}
\author{Ji\v{r}\'{i} Klime\v{s}}
\email{klimes@karlov.mff.cuni.cz}
\affiliation{Department of Chemical Physics and Optics, Faculty of Mathematics and Physics, Charles University, CZ-12116 Prague 2, Czech Republic}
\begin{document}
\maketitle
\begin{abstract}
The projector-augmented wave (PAW) method is one of the approaches that are widely used to 
approximately treat core electrons and thus to speed-up plane-wave basis set electronic structure calculations.
However, PAW involves approximations and it is thus important to understand how
they affect the results.
Tests of precision of PAW data sets often use properties of isolated atoms or of 
atomic solids.
While this is sufficient to identify problematic PAW data sets, little information has been gained
to understand the origins of the errors and suggest ways to correct them.
Here we show that interaction energies of molecular dimers are very useful not only to identify
problematic PAW data sets but also to uncover the origin of the errors.
Using dimers from the S22 and S66 test sets and other dimers we find that the error in the interaction energy 
is composed of a short range component with an exponential decay and a long range electrostatic
part caused by error in the total charge density.
We propose and evaluate a simple improvable scheme to correct the long range error 
and find that even in its simple and readily usable form, it is able to reduce the interaction energy
errors to less than one half on average for hydrogen bonded dimers.

\end{abstract}

\section{Introduction}

There are several choices one needs to make when performing total energy calculations
at the level of quantum mechanics that affect the quality of result.
We often focus on the accuracy of approximations involving Hamiltonian or wavefunction 
such as Hartree-Fock, or density functional theory (DFT) approximations.
However, there are also many numerical parameters that need to be set 
or approximations that one might use and that affect the energy as well.
When using plane-wave (PW) basis set and periodic boundary conditions (PBC)
one often approximates core electrons, either by using pseudopotentials 
(PP)\cite{hamann1979,troullier1991,vanderbilt1990} 
or the projector-augmented wave (PAW) method.\cite{Blochl2005,kresse1999}
The use of PPs or PAW reduces the size of the PW basis set needed to
describe the states and thus can speed-up the calculations.
An important parameter of PPs or PAWs is the cut-off distance from the 
nucleus in which the pseudopotential starts to differ from the all-electron potential.
The precision of the results generally decreases with increasing cut-off distance
but the pseudostates are smoother and a smaller PW cut-off can be used.
Often, there are several PPs or PAWs available that one can choose from.
For example, in the Vienna ab-initio simulations package (VASP)\cite{vasp:prb_1993, vasp:1996} code there are
Soft, Standard, and Hard PAWs available for elements from the first and second row 
of periodic table with the Standard ones being most commonly used.

The precision of different PPs or PAW data sets can be assessed by comparison 
to all-electron results calculated, e.g., with the full potential linearized plane wave 
scheme (FLAPW)\cite{FLAPW} or using Gaussian\cite{Hill_gaussian} or Slater basis sets.\cite{jensen:atomic_basis_review_2013}
For example, Lejaeghere~{\it et al.}\cite{lejaeghere2016} used the FLAPW scheme 
with local orbitals to test a large number of PPs and PAW data sets for the 
prediction of equilibrium volume and bulk modulus of elemental solids.\cite{lejaeghere2014}
However, assessments of precision of PPs or PAWs covering majority of elements are scarce 
and more often the information about precision of PPs or PAWs is only a by-product of 
study that focuses on other properties.
In this regard, Paier~{\it et al.}\cite{Kresse:PBE_PBE0_G2_1_test_jcp_2005} compared atomisation
energies for the G2-1 data set\cite{pople1989,curtiss1990}
obtained with Perdew-Burke-Ernzerhof (PBE)\cite{perdew:pbe_prl} and PBE0\cite{pbe0} 
exchange-correlation (XC) functionals in the PAW formalism to the corresponding results in 
large Gaussian basis sets.\cite{dunning:basis_set,dunning:d_func_basis_set}
Furthermore, Maggio {\it et al.}\cite{kresse:gw_jctc_2017} tested the PAW approximation
for predicting $G_0W_0$\cite{Hedin:gw_PhysRev} energies on the $GW$100 data set.\cite{setten2015}
Finally, using all-electron Gaussian reference Adllan and Dal Costro\cite{adllan2011} compared PAW and PP
schemes on a set of atomisation energies, bond distances, and vibrational frequencies
of diatomic molecules of 13 elements.
In these tests it was noted that the agreement between PW-PAW calculations
and the all-electron reference decreases for molecules containing nitrogen, oxygen, fluorine,
or sulphur. 
This is consistent with the results obtained by Lejaeghere and co-workers\cite{lejaeghere2016} but 
generally there is not much information gained than that for specific elements (oxygen and nitrogen)
harder PPs or PAWs need to be used.

Binding energies of molecular dimers or molecular solids are often affected considerably
by the choice of the PP or PAW data set.
For example, for $\alpha$ polymorph of oxalic acid one obtains a binding energy of $-1289$~meV 
using VASP's Standard PAWs when using the PBE+vdW$^{\rm TS}$ method.\cite{Tkatchenk:TS-HI}
This is several percent away from the value of $-1249$~meV obtained for the Hard PAWs.\cite{jiri2019pq_poster}
The  discrepancy between the results of Standard and Hard PAWs  can be less problematic for other systems. 
For example, the binding energy changes only by around 4~meV for ammonia crystal, from $-459$ to $-463$~meV.
Nevertheless, these imprecisions clearly affect the results of benchmarking studies.
Clearly, the interaction or binding energies are sensitive to the choice of PPs or PAWs
and can be thus used to assess their precision.
This approach was already used by Witte and co-workers\cite{witte2019} who compared the precision of 
norm-conserving\cite{hamann1979} (NC) Troullier-Martins 
pseudopotentials\cite{troullier1991} against all electron data using PBE 
and  Slater exchange with PW92 correlation (SPW92)\cite{spw92_functional} 
functionals on the S22 database.\cite{jurecka2006}
The authors noticed that the convergence of the interaction energy with the PW basis set size
is slower for molecules containing oxygen or nitrogen.
Moreover, Tosoni {\it et al.}\cite{Civalleri:jcp_2007} compared Troullier-Martins psudopotentials
to all electron calculations for molecular crystal of formic acid.
While the agreement between both approaches was considered good, the identified deviations
could be both due to unconverged Gaussian basis sets or imprecise PPs.

In this study we analyse the precision of the PAW potentials 
using interaction energies of dimers in the S22 and S66 sets\cite{jurecka2006,
S66_database} as well as some simple support dimers.
These sets are targeted on biomolecules and thus the molecules are formed by hydrogen,
carbon, nitrogen, and oxygen.
The sets cover a range of interactions, from those with large electrostatics 
contributions to those with dominant dispersion.
We test various PAW potentials supplied with the Vienna ab-initio simulation 
package (VASP)\cite{vasp:prb_1993, vasp:1996} against reference interaction energies
obtained using large Gaussian basis sets.
We obtain data for the PBE functional, which was used to construct the PAW data sets,
as well as for the HF method, which allows to assess transferability of the PAWs.
The tests on S22 and S66 are followed by calculation and analysis of errors for
binding curves of molecular dimers.
Finally, the data show that the error in the interaction energy mostly occurs due 
to incorrect description of the electron density and we devise and test 
a simple correction for the error.

\section{Methods Overview}

\subsection{{\sc VASP} calculations}

The plane-wave PAW calculations were performed using the Vienna ab-initio simulation package 
(VASP) version 6.2.\cite{vasp:prb_1993,kresse1996,kresse1999}
We obtained results for the Perdew-Becke-Ernzerhof (PBE) exchange-correlation functional\cite{perdew:pbe_prl}
and for the Hartree-Fock (HF) method.\cite{Kresse:PBE_PBE0_G2_1_test_jcp_2005}
The PBE-based Hard, Hard\_GW, Standard, Standard\_GW, and Soft potentials 
(available in \texttt{potpaw\_PBE.52} VASP PAW dataset version) were used.
These are denoted by the suffices ``{\tt \_h}", ``{\tt \_h\_GW}", none, ``{\tt \_GW}", and ``{\tt \_s}", 
respectively, in the PAW directory distributed with VASP and their properties
are summarised in Tables ST1-ST5 in the {\it Supporting Information} (SI).

The interaction energies of dimers, $E_{\rm int}$, were obtained as
$$ E_{\rm int}=E_{\rm dimer} - E_{\rm mono1} - E_{\rm mono2}\,,$$
where $E_{\rm dimer}$, $E_{\rm mono1}$, and   $E_{\rm mono2}$ are the 
energies of dimer, and the two monomers, respectively.
The error of the interaction energy with respect to reference value $E_{\rm int}^{\rm ref}$
is then
$$\Delta E=E_{\rm int}-E_{\rm int}^{\rm ref} \,.$$
For the tests of binding energies of dimers in the S22 and S66 databases
the structures were used without any structural optimisation.\cite{begdb_website}
Note that the structures of the monomers in the dimer and as isolated molecules
are identical, for this reason we use the term ``interaction energy".

Unless noted, we used plane-wave basis-set cut-offs of 2000, 1600, and 1000~eV for the Hard 
(and \texttt{h\_GW}), Standard (and \texttt{\_GW}), and Soft PAW potentials, 
respectively.
In order to minimize the interactions between periodic images, a large simulation 
cell with a side of 35~{\AA} was used to gather the data for the S22 and S66 datasets
and to perform distance scans.
Residual spurious interactions between periodic images were further reduced 
by using the dipole correction in VASP (\texttt{IDIPOL=4} tag in 
\texttt{INCAR}).\cite{Makov_Payne:PBC_dipole_correction}
The $k$-point sampling was performed only at the $\Gamma$-point.
To avoid accidental partial occupancies of one-electron states we used Gaussian smearing 
(\texttt{ISMEAR=0}) with a small smearing width of 0.01~eV (\texttt{SIGMA} tag).
The tag \texttt{LASPH=.TRUE.} was used to account for the aspherical contributions 
to the electrostatic energy within the PAW sphere.
The {\tt HFRCUT} tag was not used for the HF calculations as we are using the
same simulation cells for the dimer and monomers and thus the interaction energy
is not affected by error due to Coulomb singularity.\cite{gygi1986,jiri:elec_stru_pbc_mbe}

To help with the analysis of errors we obtained
the Bader\cite{Bader:book} and iterative Hirshfeld charges.
The iterative Hirshfeld charges were calculated within VASP by algorithm proposed 
by Bultinck~{\it et al.}\cite{bultinck:hirshfeld_algorithm, BULTINCK:cpl_2007}
as implemented by Bu\v{c}ko and co-workers 
(\texttt{IVDW=21} tag in \texttt{INCAR}.\cite{Tkatchenk:TS-HI,Angyan:ts_hi_explanation_jctc_2013})
The Bader charges were obtained using the {\sc Bader charge analysis} code
on the approximate all-electron density printed by VASP.\cite{bader_code:Henkelman_2006, bader_code:henkelman_2007,bader_code:henkelman_2009,Bader_code:Trinkle_2011}

\subsection{All electron calculations}

The  all-electron calculations of interaction energies were performed using the
{\sc Turbomole}\cite{TURBOMOLE} package, without employing the resolution-of-identity
(RI) approximation to reduce numerical errors.\cite{density_fitting}
To assure a high precision of the results, the wave function and density 
convergence criteria were set to 1$\times$10$^{-8}$~a.u., 
and the finest grid (\texttt{7} in \texttt{define}) was chosen for PBE.
The interaction energies were obtained using the Dunning's 
aug-cc-pV$N$Z ($N =$ D, T, Q, 5, and 6) basis sets, we denote them by abbreviation AV$N$Z.\cite{dunning:basis_set,dunning:d_func_basis_set}
When calculating the interaction energy for a specific dimer, all the calculations
used the dimer basis set.\cite{CP:boys_bernardi}
The data obtained with the AV5Z basis set was used as a reference for the S22 and S66 datasets.
We note that our PBE/AV5Z interaction energies for the S22 dimers are in a very good agreement
with the results obtained using a large pc-4 basis set\cite{Jensen:pc_basis_1,Jensen:pc_basis_2} by Witte and co-workers,\cite{witte2019}
the RMSE is only 0.14~meV.

To help the analysis as well as to obtain precise categorization of the considered complexes, 
we used the symmetry-adapted perturbation theory (SAPT)\cite{sapt:chem_rev} 
results of He{\ss}elmann for the S22 and S66 datasets.\cite{hesselmann:sapt_exx_jctc_2018}
Specifically, the first-order electrostatic 
($\mathrm{E^{(1)}_{elst}}$) and second-order dispersion ($\mathrm{E^{(2)}_{disp}}$) 
interaction energy components which were obtained by
localized asymptotically corrected PBE0 XC potential
(LPBE0AC)\cite{hesselmann:jcp_2005,perdew:pbe_prl} and the exact-exchange KS response 
kernel\cite{gorling:ijqc_exx_98,gorling:exx_pra_98} (EXX).

Finally, for analysis, atom-atom interaction energies were obtained by using interacting 
quantum atom  (IQA)\cite{IQA_1,IQA_2} scheme applied to PBE/AVTZ and HF/AVTZ densities.
In the IQA scheme, the dimer interaction energy is written as a sum of atom-atom 
interaction energy contributions as follows:
\begin{equation}
 \label{eqn:iqa}
 E^{IQA}_{int} = \sum^{mono 1}_{A} \sum^{mono 2}_{B}  E^{AB}_{int} = \sum_{AB}^{\mkern30mu \prime} E^{AB}_{en} + E^{AB}_{ne} + E^{AB}_{ee} + E^{AB}_{nn}\,.
\end{equation}
where A and B represent atomic basins belonging to the monomer 1 and monomer 2, respectively, 
as defined by the quantum theory of atoms in molecules (QTAIM).\cite{Bader:book}
The $E^{AB}_{int}$ is decomposed 
into nuclear-electron ($E^{AB}_{ne}$  and $E^{AB}_{en}$), electron-electron 
($E^{AB}_{ee}$), and nuclear-nuclear ($E^{AB}_{nn}$) contributions.
The $E^{AB}_{ee}$ is further divided into a sum of Coulomb ($E^{AB}_{C}$) and XC ($E^{AB}_{XC}$) contributions.
Therefore, the $E^{IQA}_{int}$ components can be categorized as:
\begin{align}
 \label{eqn:IQA_2}
 E^{IQA}_{int} = \sum_{AB}^{\mkern30mu \prime} (E^{AB}_{en} + E^{AB}_{ne} + E^{AB}_{C} + E^{AB}_{nn}) + E^{AB}_{XC} 
 = \sum_{AB}^{\mkern30mu \prime} E^{AB}_{elstat} + E^{AB}_{XC}\,. 
\end{align}
where $E^{AB}_{elstat}$ and $E^{AB}_{XC}$ are the classical electrostatic 
and XC contributions to atom-atom interactions, respectively.
All the PBE/AVTZ and HF/AVTZ densities for the IQA calculations were obtained by 
{\sc Molpro}\cite{molpro_1,molpro_2}, and the IQA calculations were carried out using {\sc AIMAll}.\cite{AIMAll} 

\section{Results}

\subsection{S22 and S66 databases}

\begin{figure}[H]
\includegraphics[width=0.7\textwidth]{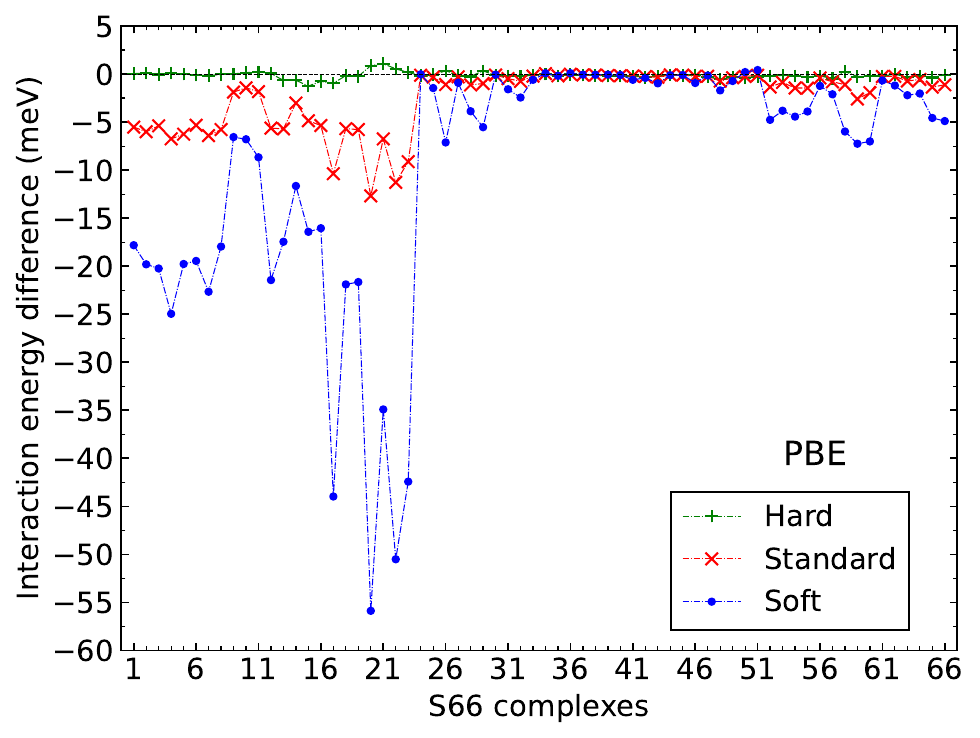}
\caption{Errors of Hard, Standard, and Soft PAW potentials for PBE interaction 
energies of S66 dimers.
\label{fig:s66_err}}
\end{figure}

We start with the errors of the different PAWs calculated for the PBE interaction
energies of the S66 dimers, shown in Fig.~\ref{fig:s66_err} (the errors
are also listed in Table ST6).
As expected, the errors are the largest in the absolute value for the Soft PAWs 
and decrease when going to Standard and Hard PAWs.
The Hard PAWs produce results of essentially reference quality, the maximum deviation from 
the reference AV5Z result is {\it ca.} $-1.2$~meV (see Table ST7).
The Standard and Soft PAWs have negligible errors only for the complexes 35--51, 
which are dispersion dominated.
The largest errors of Standard and Soft potentials can be seen for 
the hydrogen-bonded complexes, {\it i.e.}, complexes 1--23, 
the errors of the ``mixed" complexes are smaller, but still significant.
The results of the {\tt \_GW} potentials are similar to the non-{\tt \_GW} variants
and we thus do not include them in Fig.~\ref{fig:s66_err} (see Fig. SF1 and Fig. SF2 for the variation of errors of Hard$\_$GW and Standard$\_$GW across
S66 dataset, respectively).
The results obtained for the S22 database (see Tables ST8 and ST9) are qualitatively similar and we thus show them only in
the SI (Fig. SF3-SF6).
Before analysing the results in more detail we comment on the errors observed for the 
HF method and for Gaussian basis sets.

The errors obtained for the HF method (Tables ST10 and ST11 for
S66 and S22, respectively) are very similar to those of PBE. 
The individual errors are shown in Fig.~SF7 in the SI and in Table~\ref{tab:s66_stat}
we list the root mean square error (RMSE), mean absolute error (MAE), and mean error (MAE) 
on the whole S66 (see Tables ST9 and ST12 for S22 dataset).
The small differences between HF and PBE can be illustrated by the overall RMSE for Standard potential
that is {\em ca.} 4.0 and 4.4~meV for PBE and HF, respectively.
The largest difference of PBE and HF errors for Standard PAW occurs for Uracil dimer (number 17), 
where PBE and HF give errors of $-$10.5 and $-$13~meV, respectively.
For the Soft PAW, the largest difference between the PBE and HF errors is 8.7~meV for acetic acid dimer (number 20),
with the errors being $-$55.9 and $-$47.2~meV for PBE and HF, respectively.
Despite the similarities there are noticeable differences between the 
{\tt \_GW} and  non-{\tt \_GW} PAWs.
For example, the statistical errors are almost identical for PBE and HF when 
Hard{\tt \_GW} PAW is used (Table~\ref{tab:s66_stat}).
In contrast, the errors somewhat increase for the Hard PAW when PBE is replaced by HF.
Overall, the data show that the main cause of the error is likely identical for
PBE and HF calculations.
Moreover, 
the {\tt \_GW} PAWs are likely more reliable when using other functional than PBE.

In Table~\ref{tab:s66_stat} we also list the statistical errors obtained for the Gaussian
basis sets from the AV$N$Z family, the individual values are tabulated in Tables~ST6 and ST10
for PBE and HF methods, respectively.
The data show that the errors of the AVQZ basis set are marginal, this is expected
due to the exponential convergence of the total energy.
The Hard and Hard$\_$GW potentials give overall errors between that of the AVTZ and AVQZ basis sets.
For the Hard potential the errors are similar to those of the AVTZ basis set when HF energies
are calculated.
In terms of statistics, the Standard and Standard$\_$GW lead to higher average errors compared to
the counterpoise-corrected AVDZ values for the S66 database complexes.
However, the situation is not so simple as the errors vary a lot between the different dimers
and in the following we discuss how the errors depend based on the character of the binding.

\begin{center}
\begin{table}[H]
\caption{Root mean square error (RMSE), mean absolute error (MAE) and mean error (ME) 
of Hard, Hard$\_$GW, Standard, Standard$\_$GW and Soft potentials as well as 
AVDZ, AVTZ, and AVQZ basis sets with respect to AV5Z basis set on S66 data for PBE 
functional and HF method. The errors are in meV.}
\label{tab:s66_stat}
\begin{tabular}{l ccccccc}
 \hline
&\multicolumn{3}{c}{PBE}&&\multicolumn{3}{c}{HF} \\
\cline{2-4} \cline{6-8}

potential/basis set & RMSE & MAE &ME & & RMSE& MAE& ME  \\
\cline{1-4} \cline{6-8}
Hard           &0.4 &0.3&$-$0.1  &&0.9 &0.7&$-$0.7\\
Hard$\_$GW     &0.5 &0.4&$-$0.1  &&0.5 &0.4&$-$0.2\\
Standard       &4.0 &2.5&$-$2.5  &&4.4 &2.6&$-$2.6\\
Standard$\_$GW &4.0 &2.5&$-$2.5  &&4.3 &2.7&$-$2.7\\
Soft           &16.0&9.5&$-$9.5  &&14.8&8.7&$-$8.6\\
AVDZ           &3.2 &2.1&1.7     &&3.5 &1.9&1.4   \\
AVTZ           &1.4 &0.9&0.8     &&1.2 &0.7&0.7   \\
AVQZ           &0.1 &0.1&0.0     &&0.1 &0.0&0.0   \\
\hline
\end{tabular}
\end{table}
\end{center}

Let us now analyze the results in a more detail by considering statistical errors for different  
subsets of S66.
The subsets that we use are hydrogen bonded (HB), $\pi$-$\pi$, $\pi$-$\sigma$, $\sigma$-$\sigma$, and ``others''.
With few exceptions the HB and others correspond to the electrostatic and mixed groups of \v{R}ez\'{a}\v{c} {\it et al.},\cite{S66_database}
respectively, and the groups involving $\sigma$ and $\pi$ bonding introduce more fine grained classification 
of the dispersion dominated dimers.
Our classification for the individual dimers along with the original classification from Ref.~\cite{S66_database}
is given in Table~ST6.

\begin{figure}[H]
\includegraphics[width=0.7\textwidth]{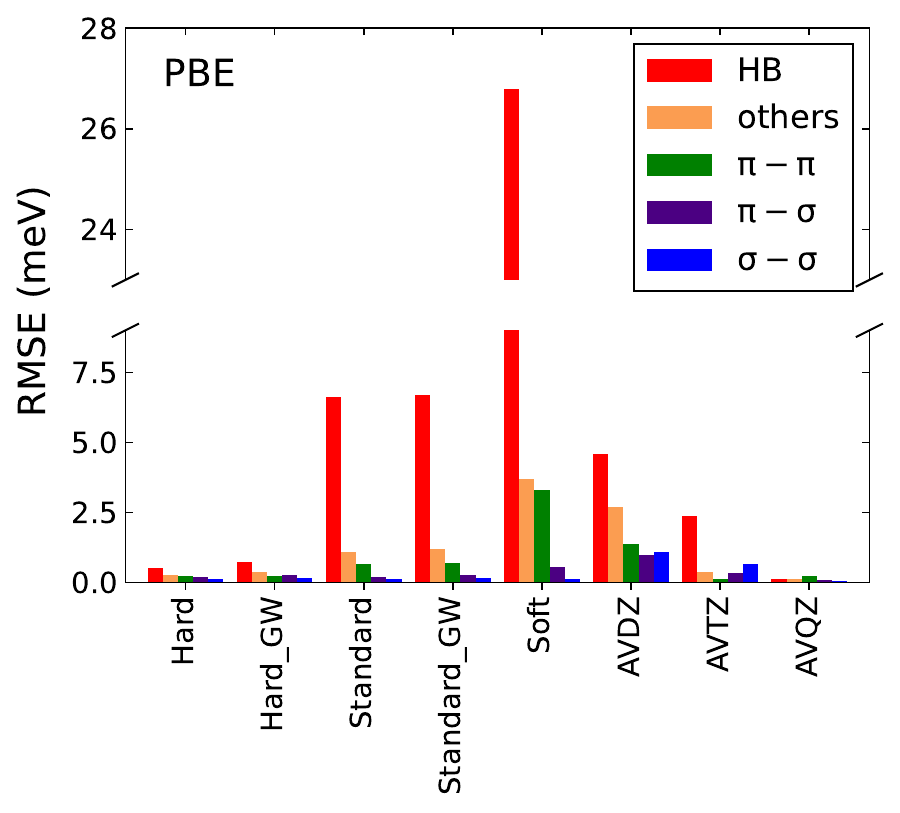}
\caption{
The root mean square error (RMSE) in meV of Hard, Hard$\_$GW, 
Standard, Standard$\_$GW,  and Soft potentials as well as AVDZ, AVTZ, and AVQZ basis 
sets with respect to AV5Z on S66 data set for different subsets of the S66 dataset.
}
\label{fig:geomstat_s66}
\end{figure}

The RMSEs of the different PAW potentials and the AV$N$Z basis sets for the different
subsets of S66 are shown in Fig.~\ref{fig:geomstat_s66} and tabulated in Table~ST13 of the SI.
All the PAWs show a similar trend: the largest RMSE is observed for the HB complexes
followed by the others, $\pi-\pi$, $\pi-\sigma$, and $\sigma-\sigma$ groups.
The errors for the HB subset are small and comparable to the errors of the other subsets 
only for the two hard PAWs.
For example, for the Hard PAW the RMSE for the HB subset is around 0.5~meV while
the RMSEs for the $\pi-\pi$ and $\pi-\sigma$ subsets are close to 0.2~meV.
The errors for the HB subset clearly dominate for the Standard and Soft PAWs,
they are at least five-times larger than the RMSEs for the other subsets.
Interestingly, the errors for the $\sigma-\sigma$ subset are small for all the PAWs, 
with RMSEs between 0.1 and 0.2~meV.

\begin{figure}
\includegraphics[width=.7\linewidth]{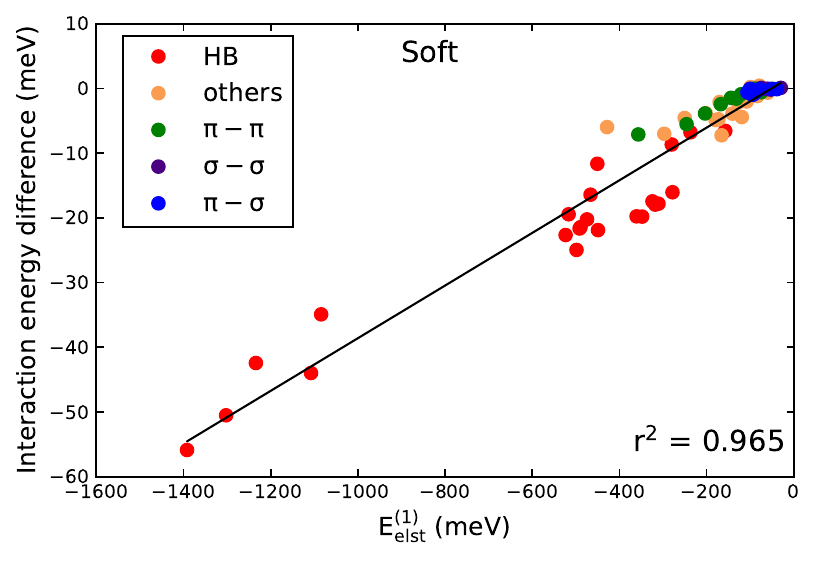}
\caption{The dependence of the errors of Soft potential
with respect to the PBE/AV5Z reference on the magnitude of the first-order electrostatic 
($E^{(1)}_{\rm elst}$) term obtained with the DFT-SAPT-EXX method for S66 database.\cite{hesselmann:jcp_2005}
The points are colored according to the subset of S66 that they belong to.}
\label{fig:corr_err_elst}
\end{figure}

The errors of the interaction energies for standard and soft PAWs increase when 
going from the $\sigma-\sigma$ to the HB subsets.
This strongly suggests that the errors depend on the 
electrostatic contribution to the interaction energy.
In order to quantitatively assess this relation we compared the PAW errors
to interaction energy components as obtained by SAPT.
This analysis shows a clear correlation between the PAW errors
and the electrostatic component of the interaction energy, 
as illustrated in Fig.~\ref{fig:corr_err_elst} for the Soft PAW.
We do not observe any significant correlations between the errors
and other SAPT components.

The statistical quantities in Table~\ref{tab:s66_stat} show that the errors of the Standard PAW 
are about 25~{\%} larger than those of the AVDZ basis set.
However, comparing the RMSEs for the different subsets one can see that only 
the RMSE of the HB subset is larger for Standard PAW than for the AVDZ basis set (Fig.~\ref{fig:geomstat_s66}).
The Standard PAW leads to smaller RMSE for the rest of the subsets.
In fact, for the $\sigma$-$\sigma$ and  $\sigma$-$\pi$ subsets the errors of the AVDZ 
basis set are even larger than those of the Soft PAW.
Therefore, the errors of the AVDZ and other Gaussian basis sets are distributed more evenly 
than the PAW errors.
As discussed by Witte~{\it et al.}\cite{witte2016} the errors for the Gaussian basis sets
tend to depend on the number of atoms in contact while for the PAWs we find the dominant 
role of electrostatic component.
Clearly these two do not need to correlate with each other leading to the qualitative differences.

\begin{figure}[H]
\begin{center}
\includegraphics[width=0.7\textwidth]{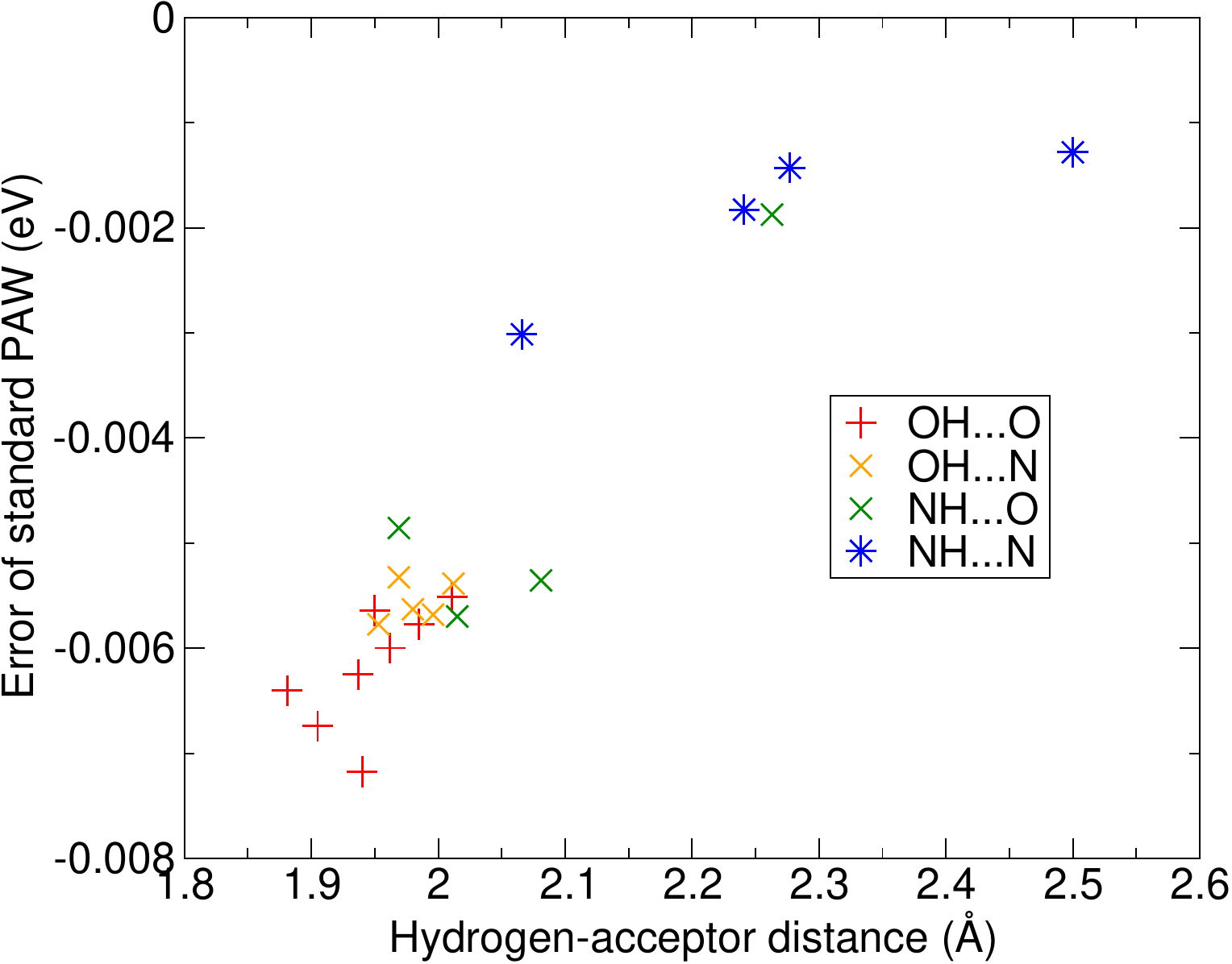}
\caption{The errors of the interaction energies of the soft PAWs for dimers
with a single hydrogen bond in the S22 and S66 datasets as a function
of the distance between the donor hydrogen and the atom accepting the hydrogen bond.}
\label{fig:erHBdist}
\end{center}
\end{figure}

As the errors are the largest for the HB dimers, we now analyse them in a more detail.
First, we take dimers with a single hydrogen bond from the S22 and S66 sets and plot
the errors for these dimers as a function of the distance between the hydrogen atom and the acceptor atom.
The data are shown in Fig.~\ref{fig:erHBdist}, divided into groups according to the 
accepting and donating atom.
One can see that the errors tend to increase in magnitude with the decrease of the hydrogen
bond length.
This is visible for all the acceptor-donor pairs but also for each group individually.
Moreover, the errors seem to be larger for the dimers involving oxygen as the donor and as the acceptor
compared to the dimers containing nitrogen as both donor and acceptor.
However, extracting more detail is difficult as the HB lengths differ for the two groups
and we return to the distance dependence of the error later.

As a final step of the analysis of the S66 data, we considered the errors of bifurcate
complexes -- dimers with two hydrogen bonds.
These are uracil-uracil, acetic acid dimer (AcOH-AcOH),
acetamide dimer (AcNH$_{2}$-AcNH$_{2}$), acetic acid-uracil(AcOH-uracil) 
and acetamide-uracil (AcNH$_{2}$-uracil), which are complexes 17, 20, 21, 22, and 23 
in  the S66 database.
As one can see in Fig.~\ref{fig:s66_err} these dimers have the largest errors.
Moreover, they also show the largest differences between HF and PBE errors.
Table~\ref{tab:bifur} lists the errors obtained for the Hard, Standard, and Soft PAWs
for the bifurcate HB complexes together with the atoms involved in the hydrogen bonds.
The errors of Soft and Standard potentials in absolute value for the bifurcate complex 
decrease in the order AcOH$\cdots$AcOH $>$ AcOH$\cdots$uracil $>$ uracil$\cdots$uracil 
$>$ AcNH2$\cdots$uracil $>$  AcNH2$\cdots$AcNH2.
As with the systems with a single hydrogen bond, the errors are clearly larger when
the HBs involve oxygen as the donor than when the donor is nitrogen.

\begin{center}
\begin{table}[H]
\caption{The error of Hard, Standard, and Soft potentials (in meV) for the bifurcate 
hydrogen bonded complexes of the S66 database as well as the atoms involved in the 
hydrogen bonds.}
\label{tab:bifur}
\small
\begin{tabular}{llccccc}
 \hline
No. &Complexes& Hard& Standard& Soft& linkage 1 & linkage 2 \\ \hline
17 & uracil-uracil     &$-$0.95 &$-$10.36 &$-$43.98 &C-O$\cdots$H-N&N-H$\cdots$O-C \\
20 & AcOH-AcOH         &0.88    &$-$12.68 &$-$55.87 &C-O$\cdots$H-O&O-H$\cdots$O-C \\
21 & AcNH$_{2}$-AcNH$_{2}$ &1.08&$-$6.75  &$-$34.91 &C-O$\cdots$H-N&N-H$\cdots$O-C \\
22 & AcOH-uracil       &0.55    &$-$11.26 &$-$50.51 &C-O$\cdots$H-N&O-H$\cdots$O-C \\
23 & AcNH$_{2}$-uracil &0.17    &$-$9.12  &$-$42.43 &C-O$\cdots$H-N&N-H$\cdots$O-C \\
\hline
\end{tabular}
\end{table}
\end{center}

\subsection{Dimer binding curves} \label{sample_system}

The data obtained for the S22 and S66 sets show a clear relation
of the PAW error to the magnitude of the electrostatic component of the interaction
and the distance between the dimers.
Moreover, the errors are the largest for molecules containing oxygen 
and nitrogen.
To get deeper insight into the errors, we constructed sets of dimers formed by
small molecules and performed distance scans of the interaction energy 
and its error.
Each set contains molecular dimers oriented in a way to create a close contact between 
two specific atoms, such as O$\cdots$O, O$\cdots$H, C$\cdots$O, H$\cdots$H, or N$\cdots$H.
Several molecules were used for each atomic pair 
to understand how the error changes when the atoms are in a different environment.
Moreover, we performed IQA analysis to corroborate the results.
In the following we only discuss the results for the O$\cdots$H and O$\cdots$O
contacts.

We start with the results obtained for the dimers involving an O$\cdots$H contact, 
i.e., a hydrogen bond involving oxygen as an acceptor.
In our set there are five dimers: a water dimer, two dimers where water is
a HB donor and the other molecules are CO$_2$ and CO with oxygen as the HB acceptor.
Finally, water acts as an HB acceptor in two dimers with the HB donating molecules
being ammonia and methane. 
The structures of all the dimers are shown in the SI in Fig.~SF8.

\begin{figure}[H]
\begin{center}
\includegraphics[width=1\textwidth]{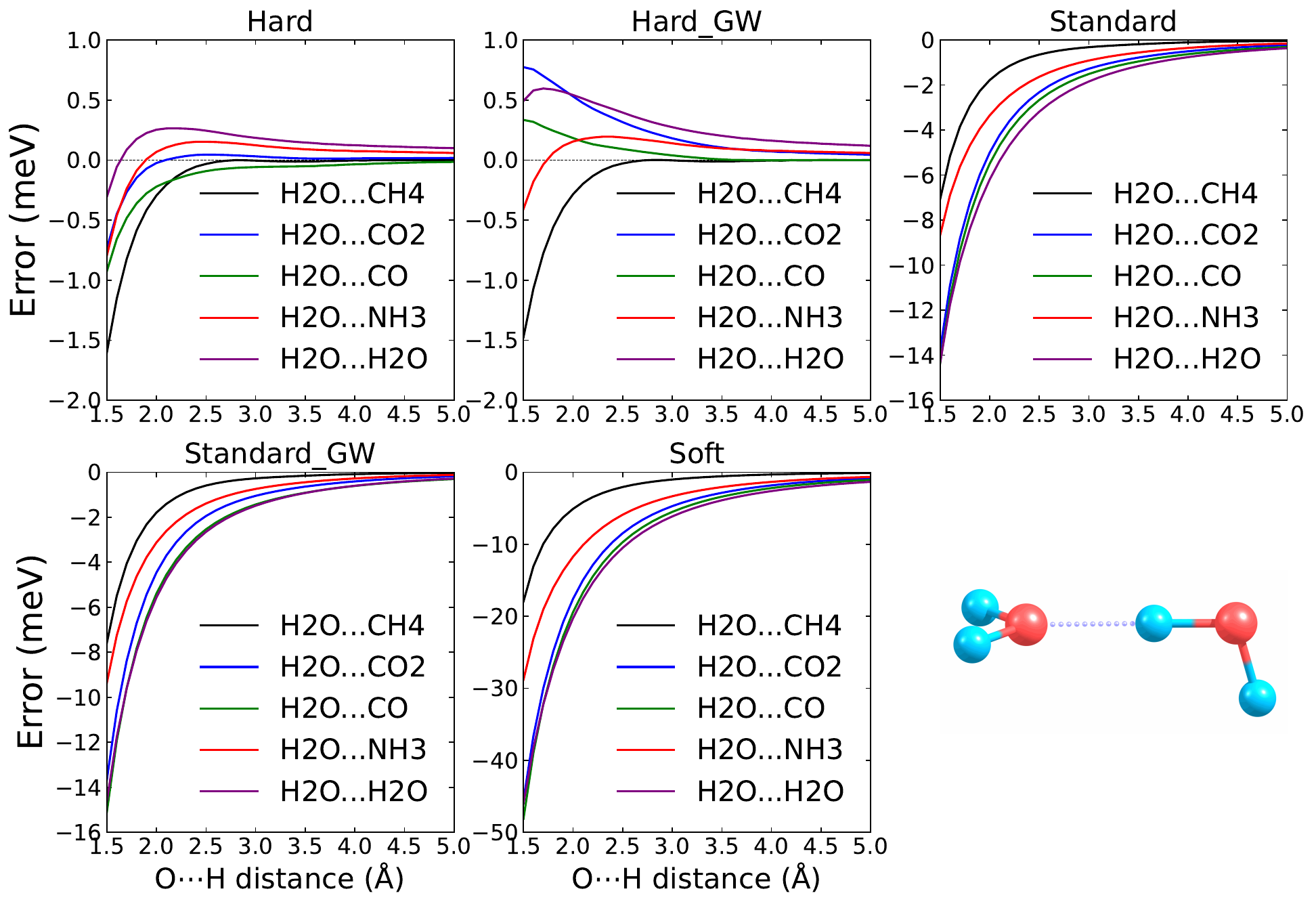}
\caption{The PBE potential energy curve errors of Hard, Hard$\_$GW, Standard, 
Standard$\_$GW and Soft potentials (with respect to PBE/AV6Z) for the 
systems with an O$\cdots$H contact in meV together with the structure of a water dimer.}
\label{fig:OHcont}
\end{center}
\end{figure}

Fig.~\ref{fig:OHcont} shows the errors of the PBE interaction energies 
for the different PAW potentials with respect to the AV6Z basis set 
as a function of distance between the bridging hydrogen and the 
HB acceptor.
As expected, the errors are again the largest for the Soft PAW and reduce 
when going to the Hard and Hard\_GW PAWs.
The errors for the two Hard PAWs are less than 0.5~meV in absolute magnitude for most of the considered 
distances and only get larger for distances below $\approx$2~{\AA}.
For either of the Hard PAWs the error seems to be a combination of a slowly
decaying component, in most cases with a positive value, and a quickly 
decreasing negative contribution at short distances. 
In contrast, the errors of the Standard and Soft PAWs are negative for 
all the distances and decay monotonously with increasing donor-acceptor distance.

We now consider the errors for the three dimers in which water accepts a HB. 
The errors clearly increase when going from methane over ammonia to water 
as the HB donating molecule.
This holds for the Standard PAWs and Soft PAW for all the distances 
and for the slowly decaying component of the Hard PAWs.
In fact, the water-methane error has apparently only the
short range component, the errors are close to zero above $\sim$3.0~{\AA}.

\begin{figure}[H]
\begin{center}
\includegraphics[width=1\textwidth]{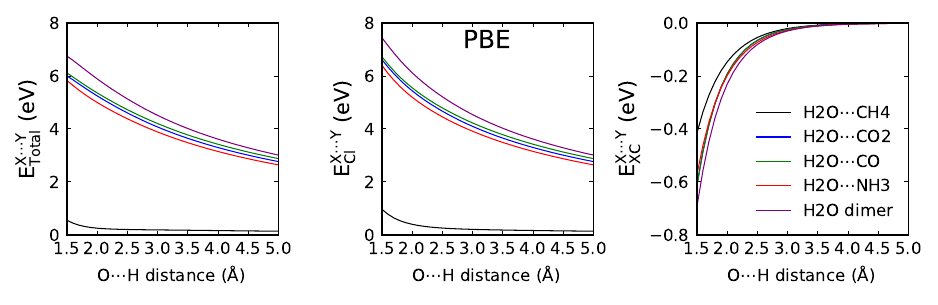}
\caption{The total X$\cdots$Y IQA interaction energy curve 
($\mathrm{E^{X\cdots Y}_{Total}}$) and its classical Coulomb 
($\mathrm{E^{X\cdots Y}_{Cl}}$) and exchange-correlation   
($\mathrm{E^{X\cdots Y}_{XC}}$) contributions for the dimers with an O$\cdots$H contact.
X and Y refers to the hydrogen atom acceptor and donor, respectively. 
All the energies are in eV.
}
\label{fig:OH_IQA}
\end{center}
\end{figure}

As the error is a combination of at least two contributions it is useful
to analyse the nature of the donor-acceptor interactions in the systems using the IQA approach.
The IQA provides information about classical electrostatic contribution to
the interaction energy and exchange correlation contribution.
The data are shown in Fig.~\ref{fig:OH_IQA} together with the 
total interaction curve between the donor and acceptor atoms.
One can see that for all the systems except methane the classical Coulomb interaction energy
is large and has a slow decay.
The exchange-correlation is much smaller and has a faster decay.
The ordering of the dimers according to the error observed for Standard PAW is 
the same as the ordering according to the magnitude of the electrostatic interaction
as provided by IQA.
Therefore, the two contributions observed in the curve for the hard PAW
can be tentatively ascribed to originate from the Coulomb interaction (long range
part) and overlap-related terms (short range). 

We now turn to the dimers containing a direct O$\cdots$O contact.
The systems are CO$_2$ dimer, O$_{2}$ dimer, CO$_{2}$ $\cdots$CO dimer, water dimer, and
complexes of water with CO and  CO$_{2}$ molecules.
The geometries are given in the SI along with the figures of the structures
in Fig.~SF9.
Note that these structures do not necessarily correspond to the lowest energy ones but such configurations
can be important nevertheless. 
For example,  water dimer with two oxygens in direct contact  appears in a high pressure 
ice VIII phase with an interatomic distance of around 3.3~{\AA}.

\begin{figure}[H]
\begin{center}
\includegraphics[width=1\textwidth]{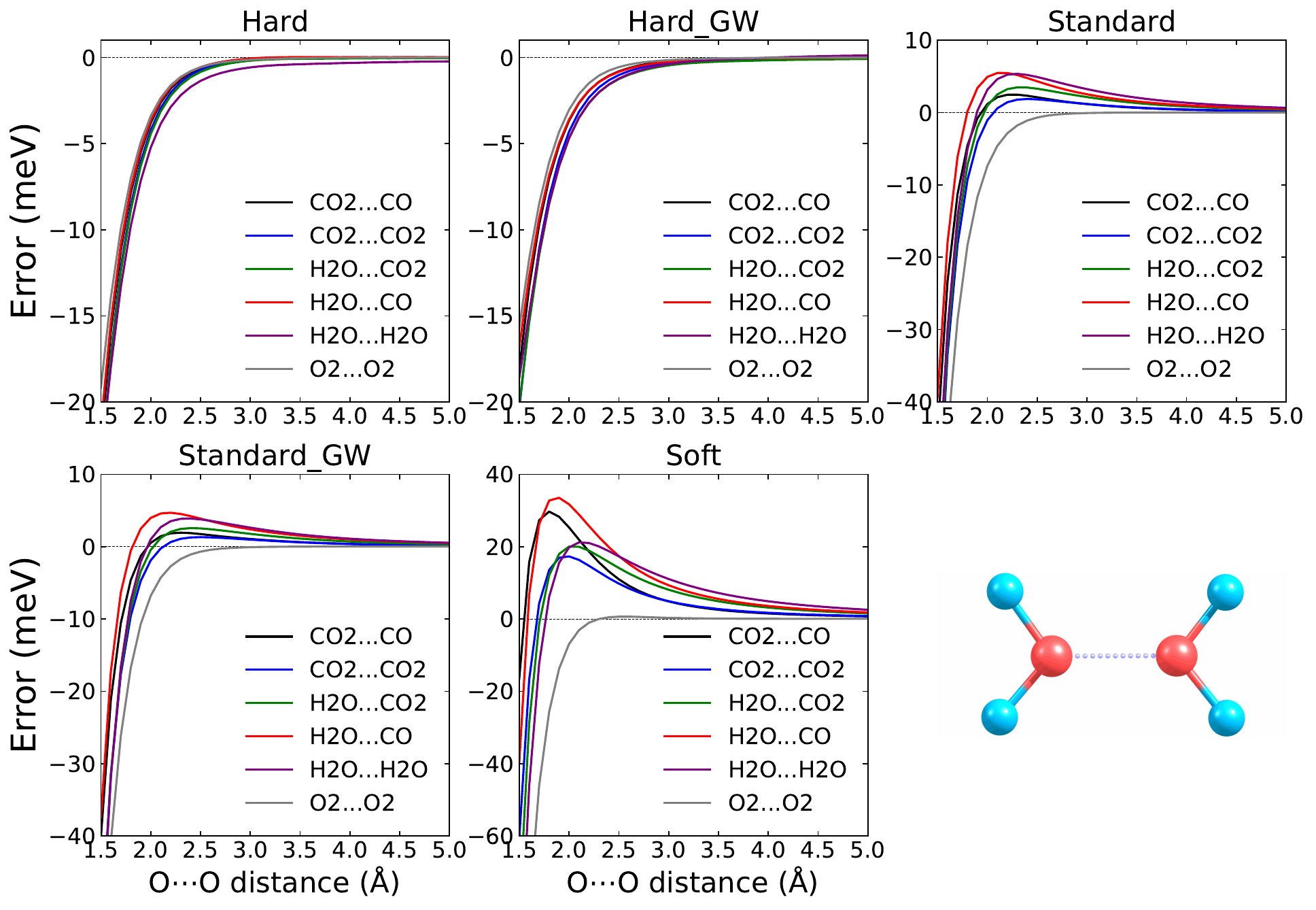}
\caption{
The PBE potential energy curve errors of Hard, Hard$\_$GW, Standard, Standard$\_$GW 
and Soft potentials (with respect to PBE/AV6Z) for the O$\cdots$O contact dimers
in meV.
Bottom right panel shows the used structure of water dimer with an O$\cdots$O direct contact.}
\label{fig:OOcont}
\end{center}
\end{figure}

The errors of the Hard, Hard\_GW, Standard, Standard\_GW, and Soft PAW potentials 
obtained for the O$\cdots$O contact systems are shown in Fig.~\ref{fig:OOcont} together with 
an image of the water dimer structure as the representive of the dimers.
As with the dimers with an O$\cdots$H contact (or hydrogen bond), two different sets of results appear.
For the Hard PAWs the error goes monotonously to zero with increasing distance between 
the molecules.
For the Standard PAWs and Soft PAW the error is negative for short distances
and positive for more distant configurations.
Note that the O$\cdots$H contact dimers showed the same characteristics, but the 
individual PAWs belonged to the opposite groups.
Therefore, the overall error is most likely composed of the two components for all the PAWs.
For the systems considered here the short range part is always negative and the long-range
one depends on the mutual orientation of the molecules so that it can be positive and negative
as discussed below.

\begin{figure}[H]
\begin{center}
\includegraphics[width=1\textwidth]{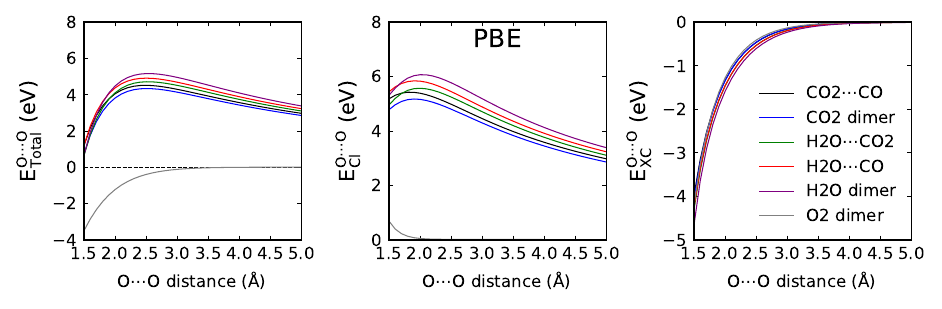}
\caption{The total O$\cdots$O IQA interaction energy curve 
($\mathrm{E^{O{\cdots}O}_{Total}}$) curve and
its classical Coulomb ($\mathrm{E^{O{\cdots}O}_{Cl}}$) and exchange-correlation 
($\mathrm{E^{O{\cdots}O}_{XC}}$)
contributions for the dimers with O$\cdots$O direct contact. 
All the energies are in eV.
\label{fig:OO_IQA}}
\end{center}
\end{figure}

We observe also considerable differences between the errors for the different molecules
forming the dimers.
The error has only the short range component for the O$_2$ dimer while the long range component
is the largest for water dimer.
The ordering of the long-range errors agrees with the ordering of the oxygen-oxygen electrostatic
interaction as provided by the IQA analysis, shown in Fig.~\ref{fig:OO_IQA}.
Clearly, the long range errors occur only for systems in which the oxygen has a non-zero 
partial charge.
When there is no zero partial charge on oxygen, like in the O$_2$ molecule, only the short
range (``overlap") component occurs.

\begin{figure}[H]
\begin{center}
\includegraphics[width=1\textwidth]{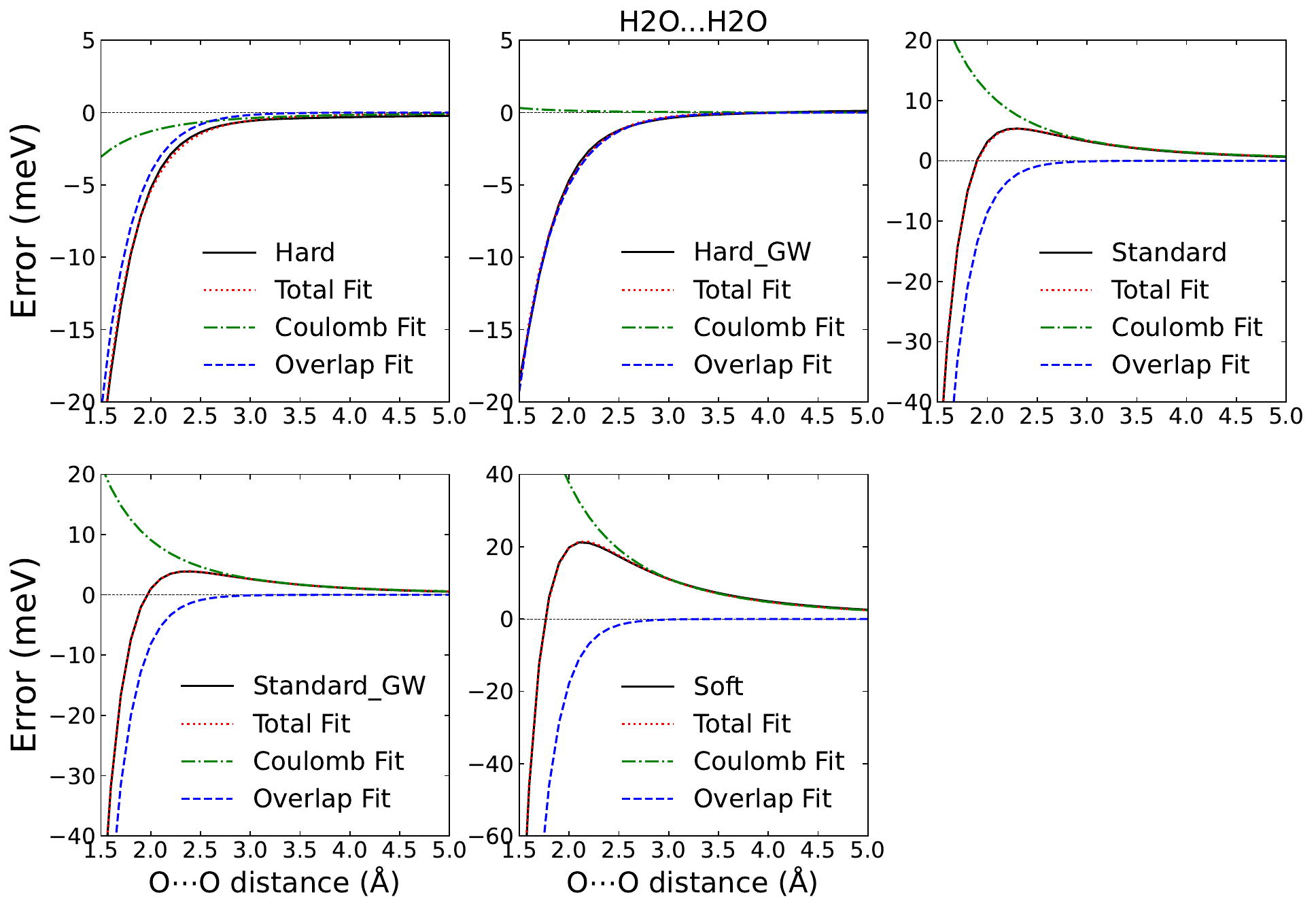}
\caption{The PBE potential energy curve errors of Hard, Hard\_GW, Standard, 
Standard\_GW and Soft potentials (with respect to PBE/AV6Z) and their fitted curves 
to $a\exp(-bR) + cR^{-3}$ (Total) as well as the Coulomb ($cR^{-3}$)
 and overlap ($a\exp(-bR)$) components 
of the errors for the O$\cdots$O contact water dimer.}
\label{fig:OO_fit}
\end{center}
\end{figure}

Overall, the binding curves show that the PAW errors have two components: a short range one 
with a fast decay that is likely related to density overlap of the molecules, 
and a second, long range one, which increases in magnitude with increasing electrostatic
component of the interaction.
Considering that the dimers are formed by neutral molecules, the leading order electrostatic
interaction is the dipole-dipole term.
A dipole-dipole interaction could occur from errors of electron density caused by the use 
of the PAW approximation.
This suggest that the error could be fitted by the function
\begin{equation}
\label{fitting}
E^{\rm err}_{\rm int}(R) = ae^{-bR} + cR^{-3}\,,
\end{equation}
in which the exponential function models the overlap component and the second term
corresponds to the dipole-dipole interaction, 
and  $a$, $b$, and $c$ are coefficients of the fit and $R$ refers to the intermolecular
or interatomic distance.

We show an example of the fit using Eq.~\ref{fitting} for water dimer with O$\cdots$O 
contact in Fig.~\ref{fig:OO_fit}.
The variable $R$ is the distance between the oxygen atoms.
The simple model fits the data very well for all the PAW potentials.
The Hard PAWs show almost negligible errors for the electrostatic component
and the exponential contribution dominates.
The magnitude of the exponential component increases when going to the Standard and Soft
PAWs.
For distances above $\sim$1.7~{\AA} the error due to the electrostatic component becomes larger
than the overlap error.
Moreover, the error clearly follows the $R^{-3}$ asymptotic behavior and thus corresponds
to incorrect description of the dipole-dipole interaction stemming from incorrect electron density.
This can be explicitly verified by calculating the dipole of water molecule using the 
different PAW potentials.
We indeed find that the dipole is 0.3747~e{\AA} for the Hard PAW but increases to 0.3801~e{\AA} for the Standard PAW and 
to 0.3914~e{\AA} for the Soft PAW.
Finally, the fact that the long-range component is caused by the dipole-dipole term explains
the changes of the errors when going from the hydrogen bonded dimers 
(Fig.~\ref{fig:OHcont}) to the O$\cdots$O contact (Fig.~\ref{fig:OOcont}).
When the water molecule is flipped, the dipole turns as well and the dipole-dipole interaction
changes sign.

\subsection{Electrostatic correction}

The long-range component of the PAW error comes from the Coulomb interaction
and the cause of the error therefore needs to be incorrect density of the monomers.
To assess the magnitude of the density errors we considered several molecules 
from the S22 and S66 test sets and obtained charge density differences ($\Delta\rho$) 
between Soft and Hard or Standard and Hard PAWs.
For this comparison we used the approximate all-electron densities printed
by VASP by setting {\tt LAECHG=.TRUE.} in INCAR.

\begin{figure}[H]
\begin{center}
\includegraphics[width=0.5\textwidth]{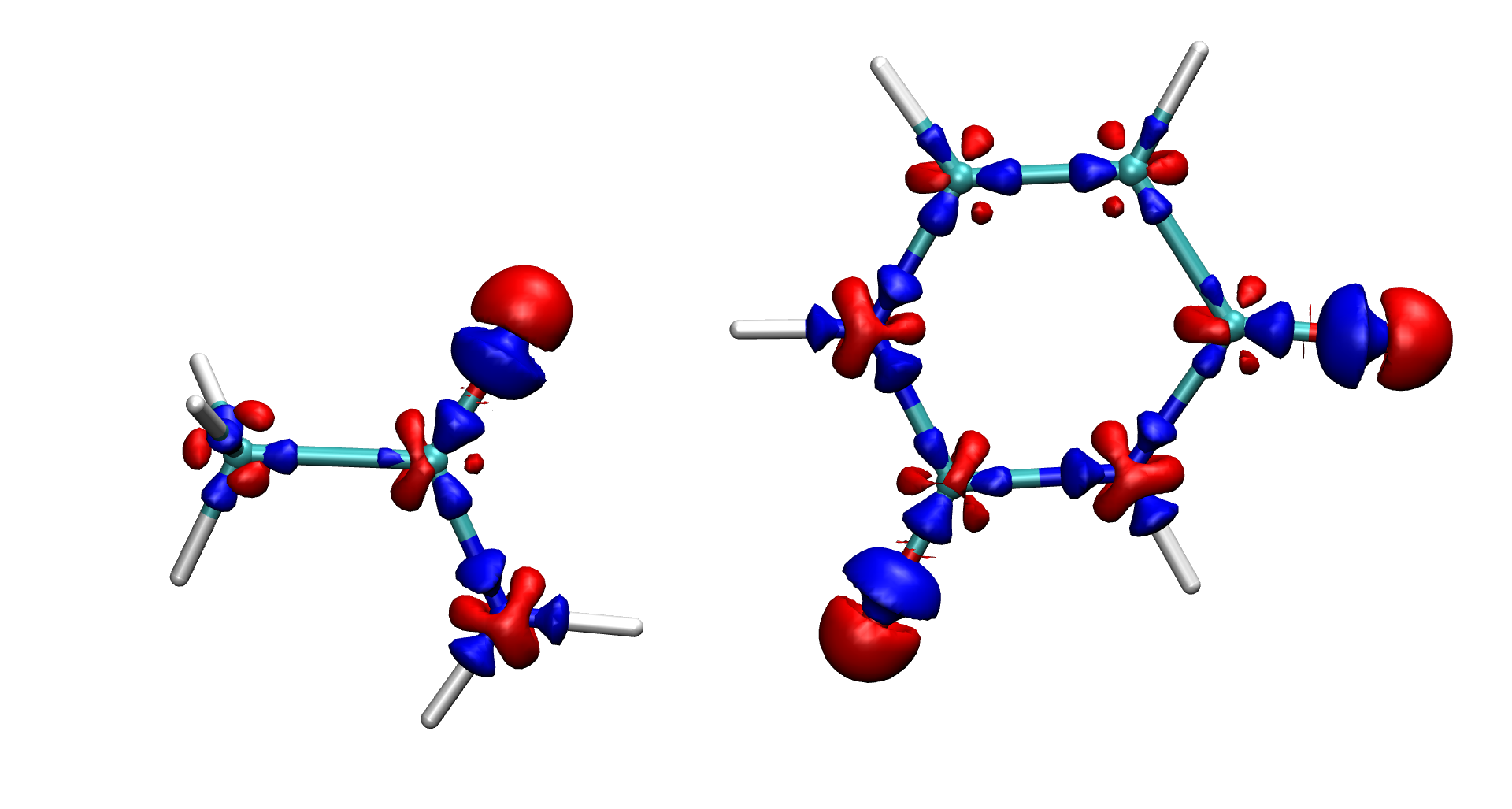}
\hskip 1em
\includegraphics[width=0.5\textwidth]{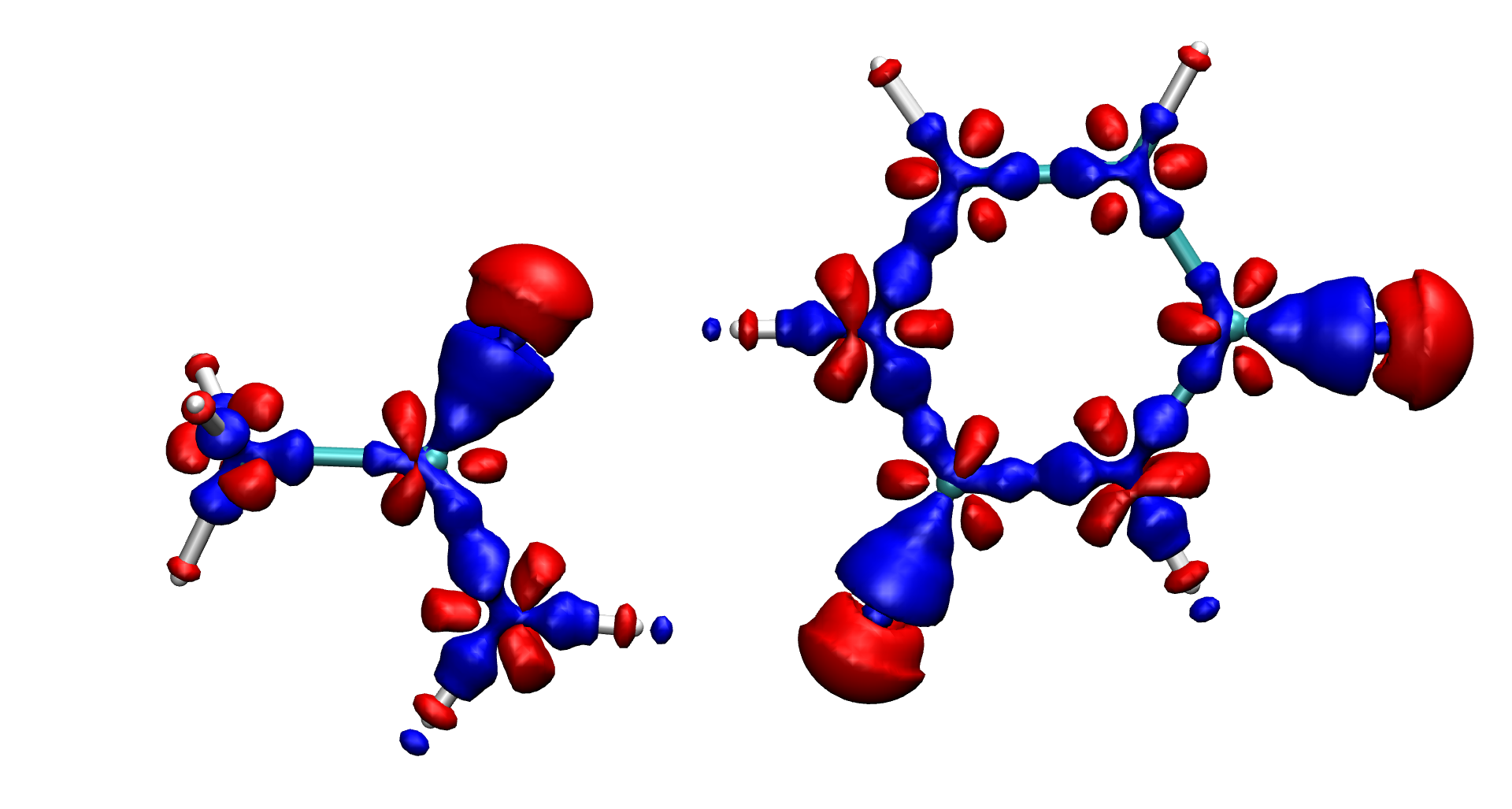}
\caption{Electron density difference between Standard and Hard PAWs (top) and 
Soft and Hard PAWs (bottom) of acetamide-uracil dimer. 
The red and blue iso-surfaces are plotted at values of $-0.02$ and 0.02~{\AA$^{-3}$}.}
\label{fig:EDD}
\end{center}
\end{figure}

Fig.~\ref{fig:EDD} shows an example of $\Delta\rho$ for acetamide-uracil dimer
for the Standard PAW (top) and Soft PAW (bottom) with the isosurfaces showing
values of $\pm0.02$~{\AA$^{-3}$}.
There is a substantial electron density difference around the oxygen atom,
the differences are smaller for carbon and nitrogen while the density differences 
around hydrogens are visible only for the Soft PAW.
The density difference for oxygen resembles a $p$-like function and thus has a dipole moment (see Fig.~SF10 as well).
This is consistent with our previous findings that the interaction energy error 
can be fitted with a dipole-dipole term, and the observation that the error correlates with the distance between donor-acceptor atoms.
We find similar electron density errors around oxygen atoms in other molecules, the density errors
are of comparable size for nitrogen atoms with two neighbors or with three neighbors
in a structure similar to ammonia (i.e., not all four atoms in plane).
When the nitrogen atom lies in the plane containing its three neighbors, as in uracil 
shown in Fig.~\ref{fig:EDD}, the overall density error lacks an out-of-plane dipole. 
The density error in the planar structure resembles a sum of three dipoles, each pointing
in a direction of a covalent bond to one of the neighbors.
The density differences for carbon atoms also resemble a sum of dipole-like density errors 
contributed by each chemical bond.
Therefore, the density errors have no or very small dipoles around carbon atoms for the
molecules in the S22 and S66 data sets.
Overall, these observations are consistent with the interaction energy errors calculated
for the S22 and S66 dimers.

We now consider the part of the interaction energy due to density-density 
electrostatic interactions and use it to derive several formulae that
can be used to quantify and correct the observed errors.
First, the electrostatic interaction between electron densities of two monomers ($\rho_{\rm A}$
and $\rho_{\rm B}$) is
\begin{equation}
E^{\rm elst}_{\rm AB}=\int\int \rho_{\rm A}({\bf r}_1) \frac{1}{|{\bf r}_1-{{\bf r}_2|}}\rho_{\rm B}({\bf r}_2) d{\bf r}_1 d{\bf r}_2\,.
\label{eq:el1}
\end{equation}

We define a density error of monomer A as 
\begin{equation}
\Delta \rho_{\rm A}=\rho_A^{\rm S}-\rho_A^{\rm H}\,,
\label{eq:el2}
\end{equation}
where  $\rho_A^{\rm S} $ and $ \rho_A^{\rm H}$ are respectively the densities obtained by more approximate
PAW potential ``S" (such as Standard or Soft) and a more precise one, such as that 
calculated using the Hard PAW potential.
The density error for monomer B is defined analogously.

The error in the density-density electrostatic component of the interaction energy is
\begin{equation}
\Delta E=E^{\rm elst, S}_{\rm AB} - E^{\rm elst, H}_{\rm AB}\,,
\label{eq:el3}
\end{equation}
the additional index compared to Eq.~\ref{eq:el1} indicates that the first energy is calculated with the approximate density 
and the second with the precise one.
Using the relations for electrostatic interaction of two electron densities  (Eq.~\ref{eq:el1})
and for the density errors of monomers A and B (Eq.~\ref{eq:el2}) in Eq.~\ref{eq:el3}
we obtain
\begin{equation} \label{eq:el4}
\begin{split}
\Delta E=\int\int \rho_{\rm A}^{\rm H}({\bf r}_1) \frac{1}{|{\bf r}_1-{\bf r}_2|}\Delta\rho_{\rm B}^{\rm S}({\bf r}_2) d{\bf r}_1 d{\bf r}_2 \\
+\int\int \Delta\rho_{\rm A}^{\rm S}({\bf r}_1) \frac{1}{|{\bf r}_1-{\bf r}_2|}\rho_{\rm B}^{\rm H}({\bf r}_2) d{\bf r}_1 d{\bf r}_2 \\
+\int\int \Delta\rho_{\rm A}^{\rm S}({\bf r}_1) \frac{1}{|{\bf r}_1-{\bf r}_2|}\Delta\rho_{\rm B}^{\rm S}({\bf r}_2) d{\bf r}_1 d{\bf r}_2\,.
\end{split}
\end{equation}
This can be further simplified if we use an expression for the electrostatic potential created by electron density
$V^{\rm H}_{\rm A}({\bf r}_2)=\int \rho_{\rm A}^{\rm H}({\bf r}_1) \frac{1}{|{\bf r}_1-{\bf r}_2|}d{\bf r}_1$
and define $V^{\rm H}_{\rm B}({\bf r}_1)$ in a similar way.
We obtain 
\begin{equation}
\Delta E=\int V^{\rm H}_{\rm A}({\bf r}_2) \Delta\rho_{\rm B}^{\rm S}({\bf r}_2) d{\bf r}_2
+\int \Delta\rho_{\rm A}^{\rm S}({\bf r}_1)  V^{\rm H}_{\rm B}({\bf r}_1) d{\bf r}_1 
+\int\int \Delta\rho_{\rm A}^{\rm S}({\bf r}_1) \frac{1}{|{\bf r}_1-{\bf r}_2|}\Delta\rho_{\rm B}^{\rm S}({\bf r}_2) d{\bf r}_1 d{\bf r}_2\,.
\label{eq:el5}
\end{equation}
The first two terms give the interaction of the density error of one monomer with the 
potential of the other monomer, the last term then describes the interaction of the
two density errors.
The  last term is likely to be smaller compared to the first two contributions.
If we neglect it we obtain
\begin{equation}
\label{eq:mod1}
\Delta E_{\rm elst} = \int\Delta\rho_{\rm A}^{\rm S}({\bf r})V^{\rm H}_{\rm B}({\bf r})d{\bf r}  +
\int\Delta\rho_{\rm B}^{\rm S}({\bf r})V_{\rm A}^{\rm H}({\bf r})d{\bf r}\,,
\end{equation}
which is the first relation that we use to compare to the actual error of different PAWs.
This equation can be readily tested for VASP as all the required quantities (electron densities and 
Hartree potential) can be written by the code and processed by an external program.

The form of the error suggests several ways how to estimate it using more approximate means.
One possibility is to approximate the electrostatic potential of a molecule by
using atom-centered partial charges as
\begin{equation}
V^{\rm approx}({\bf r})=\sum_i \frac{q_i}{|{\bf r}-{\bf r}_i|}\,,
\end{equation}
where $r_i$ are the positions of the atoms and $q_i$ are the partial charges.
Moreover, the density error can be approximately expanded into atom-centered
multipole moments.
In this work we use only the dipole moment.
Using these two approximations in Eq.~\ref{eq:el5} gives an approximate 
formula for the interaction energy error 
\begin{equation}
\label{eq:mod2}
\Delta E_{\rm dipole} = \sum_{i\in \rm A}\sum_{j\in \rm B}\left[ 
\frac{(\boldsymbol{\mu}_{i}\cdot\textbf{r}_{ij})q_{j}}{r_{ij}^{3}} +
\frac{(\boldsymbol{\mu}_{j}\cdot\textbf{r}_{ij})q_{i}}{r_{ij}^{3}} +
\frac{\boldsymbol{\mu}_{i}\cdot\boldsymbol{\mu}_{j}}{r_{ij}^3} -
3\frac{(\boldsymbol{\mu}_{i}\cdot\boldsymbol{r}_{ij})(\boldsymbol{\mu}_{j}
\cdot\boldsymbol{r}_{ij})}{r_{ij}^5}\right]\,,
\end{equation}
where the indices $i$ and $j$ run over atoms in monomers A and B,
respectively, 
$\mu_i$ are the dipoles assigned to atom $i$,
and ${\bf r}_{ij}$ is the vector between atoms $i$ and $j$.
Note that in practice we use point charges obtained for the approximate
PAW potentials and not for the more precise PAWs in Eq.~\ref{eq:mod2} as we want to avoid 
calculations with the precise PAWs.

There are several methods that can be used to obtain the atomic charges.
We have tested three of them here: Hirshfeld, iterative Hirshfeld, and Bader schemes.
The main reason for choosing these approaches is that they are accessible 
from VASP.
Initial tests of the Hirshfeld scheme showed unsatisfactory charges for water 
(too small). 
Therefore, we only discuss the results obtained with the iterative
version for which we obtained more satisfactory results.

From the elements that are present in the S22 and S66 sets (hydrogen, carbon, nitrogen, oxygen) 
the density errors are the largest 
for oxygen and nitrogen and we therefore only set the dipoles of the density error for these atoms in Eq.~\ref{eq:mod2}.
To obtain the magnitudes of the dipole moments of the density errors for oxygen and nitrogen we 
calculated the dipole moments of water, formaldehyde, and methylamine using the different PAW potentials.
Water was used to obtain dipole error for oxygen with two neighbors and formaldehyde
for oxygen with a single neighbor.
The magnitude obtained with the hard PAW was taken as reference and the 
difference for Standard and Soft PAWs was the sought dipole error magnitude.
These values are summarised in Table~\ref{tab:dips}.

The density errors resemble a dipole directed along an axis of rotation of the molecule
(if there is exactly one).
To obtain the direction of the dipole error vector we first calculate the sum of 
normalised vectors to neighboring atoms $j$
\begin{equation}
\label{eq:dip_dir}
{\bf v}_i = \sum_{j}\hat{\bf r}_{ij}\,.
\end{equation}
We then check the length of ${\bf v}_i$ as a small value indicates that atom $i$ and its neighbors 
lie approximately in the same plane and the dipole moment of the density error, at least 
for the molecules considered here, is small.
As mentioned above, this can occur for nitrogen, e.g., in uracil.
If the length of ${\bf v}$ is equal or above $0.75$, we normalize it and use it
as the dipole direction.
The final dipole error is then obtained by multiplying the direction with appropriate
magnitude.

\begin{center}
\begin{table}[H]
\caption{The atom-centered dipole moment errors used in the correction scheme.}
\label{tab:dips}
\begin{tabular}{lllc}
 \hline
Element & No. neighbors & PAW & Dipole error (e{\AA}) \\
\hline
O    &  1 & Standard & 0.0102      \\
O    &  2 & Standard & 0.0076  \\
O    &  1 & Soft & 0.0385  \\
O    &  2 & Soft & 0.0240  \\
N    &  -- & Standard & 0.0041 \\
N    &  -- & Soft & 0.0161\\
\hline
\end{tabular}
\end{table}
\end{center}

\begin{figure}[H]
\begin{center}
\includegraphics[width=1.0\textwidth]{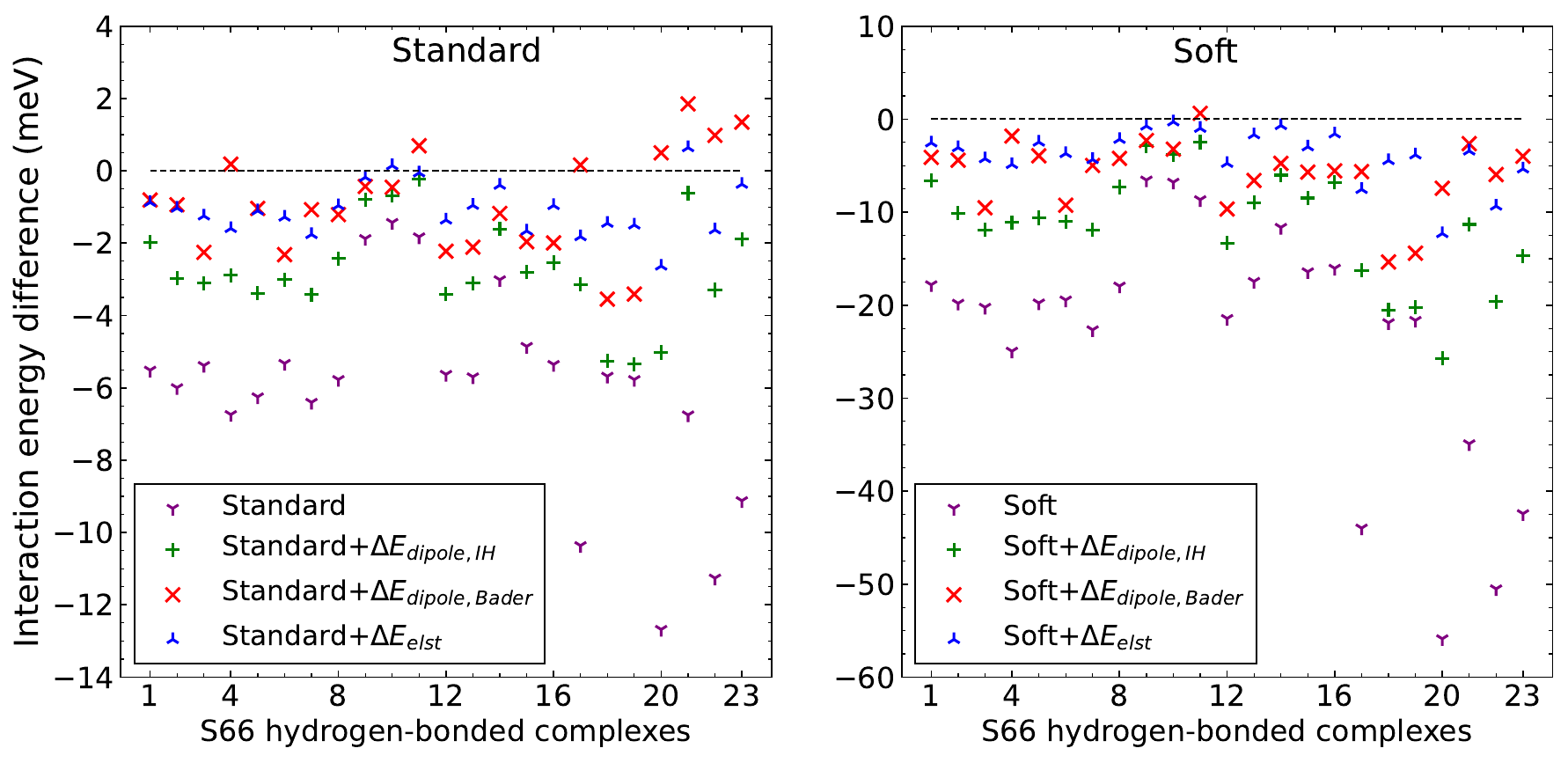}
\caption{Errors of PBE interaction energies obtained for Standard and Soft PAWs for the HB bonded complexes 
of the S66 data set.
Bare data are shown with dots and corrected for the electrostatic error using various schemes:
$\Delta E_{\rm dipole,Bader}$ and $\Delta E_{\rm dipole,IH}$ were 
calculated using Eq.~\ref{eq:mod2} with Bader and iterative Hirshfeld charges, 
respectively, while $\Delta E_{\rm elst}$ corrections were obtained by using Eq.~\ref{eq:mod1}.
}
\label{fig:elstat_cor}
\end{center}
\end{figure}

We used the models defined in Eqs.~\ref{eq:mod1} and~\ref{eq:mod2} to correct the
errors of Standard and Soft PAW potentials on the S66 data set.
The bare and corrected errors of the Standard and Soft PAW potentials are shown in Fig.~\ref{fig:elstat_cor}
for hydrogen bonded dimers and the average errors on the whole S66 set and its subsets 
are given in Table~\ref{tab:elstat_cor}.
All the corrections reduce significantly the errors of the two PAW potentials.
The lowest average errors are observed for the least approximate density-based correction, the average
errors for hydrogen bonded dimers are reduced by a factor of $\approx$6 both for the 
Standard and Soft PAWs.
The Bader-charges based correction leads to similar average errors as the density correction
for the Standard PAW, but gives larger average errors for the Soft PAWs.
One possible reason is that the density errors are larger for the Soft PAWs and 
they are less well approximated using only the dipole term employed in the Bader-based correction.
Therefore, terms beyond dipole for the charge density error and beyond monopole for the charge density
are likely needed to improve the results.
In any case the Bader-corrected Soft PAW leads to similar errors as the uncorrected 
Standard PAW, at least for the hydrogen bonded systems, the reduction of errors is less
significant for the mixed and dispersion bonded subsets.

The correction based on the iterative Hirshfeld charges is overall less efficient compared to
the Bader-based scheme.
Specifically, for the hydrogen bonded dimers the iterative Hirshfeld 
corrections are always smaller than those that use the Bader charges (Fig.~\ref{fig:elstat_cor}). 
The correction with iterative Hirshfeld charges is particularly small for dimers 18 and 19 which involve
pyridine interacting with water and methanol, respectively, see Fig.~\ref{fig:elstat_cor}.
Also in this case the iterative Hirshfeld charges are lower than the Bader ones.
The scheme utilising iterative Hirshfeld charges reduces the errors for the mixed and dispersion 
bonded groups in a similar way to the Bader-based method.

\begingroup
\begin{center}
\begin{table}[H]
\caption{The mean bare and corrected errors (in meV) with respect to AV5Z
for the Soft and Standard PAW potentials
for the S66 test set and its subsets. 
HB, M, and D refer to hydrogen bonded, mixed electrostatic-dispersion, and dispersion stabilized complexes.}
\label{tab:elstat_cor}
\begin{threeparttable}
\begin{tabular}{llcccc}
 \hline
PAW &  Correction & S66 & HB & M & D  \\
\hline
Standard  &  none$^{a}$        & $-2.5$ & $-6.0$ & $-0.9$  &  $-0.3$  \\\
          &  $\Delta E^{b}_{\rm elst}$     & $-0.6$ & $-1.0$ & $-0.4$ & $-0.3$\\\
          &  $\Delta E^{c}_{\rm dipole,Bader}$       & $-0.6$ & $-0.9$  &$-0.6$  & $-0.3$ \\\
          &  $\Delta E^{d}_{\rm dipole,IH}$& $-1.1$ & $-2.7$  & $-0.6$ & $0.0$ \\\
Soft      &  none              & $-9.4$ & $-23.4$ & $-2.9$  &  $-1.2$  \\\
          &  $\Delta E_{\rm elst}$            & $-1.7$ & $-3.8$ & $-0.8$ & $-0.5$\\\
          &  $\Delta E_{\rm dipole, Bader}$            & $-3.1$ & $-5.9$  &$-2.1$  & $-1.2$ \\\
          &  $\Delta E^{d}_{\rm dipole, IH}$  & $-4.5$ & $-11.4$  & $-1.7$ & $-0.1$ \\
\hline
\end{tabular}
\begin{tablenotes}
\item[$^{a}$] the bare PAW potential error.
\item[$^{b}$] using correction of Eq.~\ref{eq:mod1}.
\item[$^{c}$] using Eq.~\ref{eq:mod2} and Bader atomic charges.
\item[$^{d}$] using Eq.~\ref{eq:mod2} and iterative Hirshfeld atomic charges.
\end{tablenotes}
\end{threeparttable}
\end{table}
\end{center}
\endgroup

\begin{figure}
\includegraphics[width=.7\linewidth]{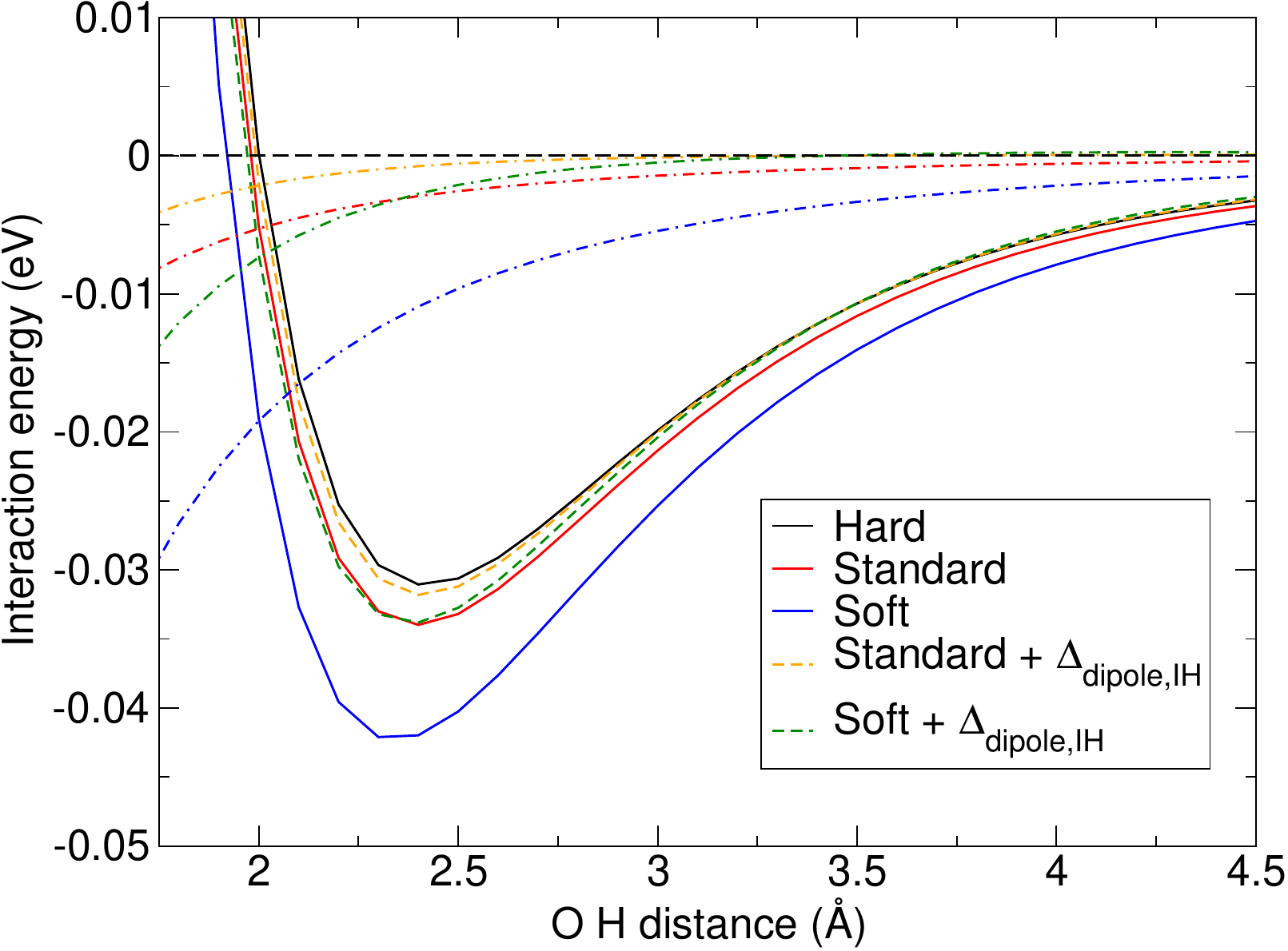}
\caption{Binding curves of water and CO dimer obtained for the Hard, Standard, 
and Soft PAW data sets (solid lines), the data with correction according
to Eq.~\ref{eq:mod2} applied to Standard and Soft data (dashed lines).
The dash-dot lines show the difference to the result obtained for the Hard
PAW which is taken as a reference.
The Iterative Hirshfeld charges calculated by VASP were used to calculate the
correction.}
\label{fig:corr_curve}
\end{figure}

To illustrate how the correction scheme performs for binding curves
we show data for water-CO dimer with O$\cdots$H contact in Fig.~\ref{fig:corr_curve}.
For the Standard and Soft PAWs we give the curves without as well as with the correction 
according to  Eq.~\ref{eq:mod2} employing iterative
Hirshfeld charges and oxygen-centered dipoles tabulated in Tab.~\ref{tab:dips}.
Moreover, we show the errors of the different settings with respect to the Hard PAWs 
by the dash-dot curves.
One can see that the corrections considerably reduce the errors of both Soft and 
Standard PAWs for all the distances.
Around the equilibrium distance (around 2.5~{\AA}), the error of the Soft PAW is reduced 
from around $-9.6$~meV to $-2.1$~meV.
Moreover, the corrected Soft PAW has clearly better asymptotic behavior compared to the
uncorrected one, the correction brings it to within 0.5~meV of the Hard PAW reference for distances
above 3.0~{\AA}.
The errors of the corrected Soft PAW are even smaller than the errors of the uncorrected
Standard PAWs for distances larger than the equilibrium distance.
Also note that the interaction energy minimum on the Soft PAW curve is shifted to smaller 
distances, by around 0.1~{\AA}, and the correction partly remedies this issue.
For the Standard PAW the error at equilibrium decreases from $-2.6$~meV to $-0.6$~meV
and the difference to the Hard data for distances above 3~{\AA} is around 0.1~meV or smaller.

\section{Summary and conclusions}

We used interaction energies of molecular dimers to test
precision of several PAW data sets supplied with the VASP code.
In general, we obtain similar qualitative results as obtained by 
previous tests\cite{lejaeghere2016,setten2015,adllan2011},
that is, we find that errors are larger for oxygen and nitrogen compared to carbon and hydrogen.
However, compared to the previous works we clearly see that
the errors originate from incorrect electron density leading
to differences in electrostatic interactions.

We identify two components of the error: a short range one with an
exponential decay and a long-range one with an algebraic decay.
The first one is less important for typical bond lengths of weakly bonded dimers,
but, of course, it will affect atomisation energies, covalent bond lengths, 
or very short hydrogen bonds.
The long-range error originates from errors in electron density that appear
for more approximate PAW potentials (Standard, Soft, \dots).
For the molecules studied here the density errors can be described as a sum of $p$-like functions
centered at a given atom with each function associated with one covalent bond.
The total density error than can be mostly $p$-like, such as for water or ammonia,
or have a higher electrostatic moment such as for nitrogen atoms in uracil or for carbon atoms.
This shape of the density error suggests a way to correct the error in the interaction energy.
The density error can be modelled by atom-centered dipoles that interact electrostatically
with the rest of the system.
Moreover, the rest of the system can be approximated by atom centered point charges.
Even such simple corrections reduce the errors to around one half for most 
of the hydrogen bonded dimers.
They are less efficient for the dispersion bonded dimers where the density
errors are typically of a higher multipole, but extension to these cases
can be done by, e.g., assigning a density error to each covalent bond.

The proposed corrections for the long-range component are simple
to implement and can be used to process an output of existing calculations or 
to correct the energies (and forces) on the fly.
The only non-trivial requirement are the point charges, even though 
the charge density of the system could be used directly as well.
Currently, we obtain Bader charges from the all-electron density produced by VASP
using an external tool.
This is less convenient compared to the calculation of the iterative Hirshfeld charges
which can be done directly in the VASP code.\cite{Bucko2013}
However, the iterative Hirshfeld charges lead to worse results compared to
the Bader charges, at least for the hydrogen bonded systems.
The short-range component, extending to some 1.0--1.5~{\AA} from nuclei, could
be modelled using exponential potentials such as used in forcefields or by 
a general machine learned forcefield.
The latter approach could  be also used to train and then predict the density errors,
in similar spirit to learning total density.\cite{thurlemann2022}

We demonstrated reduction of the error for molecular dimers but the scheme can be
easily extended to molecular solids or to molecular adsorption for which Standard PAWs are
widely employed.
Moreover, we expect that the origin of error is likely identical for pseudopotentials
and the correction could be used for them as well.
Finally, as the simple correction reduces the errors of Soft PAWs to a level of Standard PAWs
it suggests that the corrected Soft PAWs could be used for situations in which Standard PAWs are currently 
sufficiently precise.
This would reduce computational cost of calculations or increase accessible system size.

\begin{acknowledgement}
This work was supported by the European Union's Horizon 2020 research and innovation
program via the ERC grant APES (No 759721) 
and by the Ministry of Education, Youth and Sports of the Czech Republic through 
the e-INFRA CZ (ID:90140).
\end{acknowledgement}

\providecommand{\latin}[1]{#1}
\providecommand*\mcitethebibliography{\thebibliography}
\csname @ifundefined\endcsname{endmcitethebibliography}
  {\let\endmcitethebibliography\endthebibliography}{}


\begin{mcitethebibliography}{62}
\providecommand*\natexlab[1]{#1}
\providecommand*\mciteSetBstSublistMode[1]{}
\providecommand*\mciteSetBstMaxWidthForm[2]{}
\providecommand*\mciteBstWouldAddEndPuncttrue
  {\def\EndOfBibitem{\unskip.}}
\providecommand*\mciteBstWouldAddEndPunctfalse
  {\let\EndOfBibitem\relax}
\providecommand*\mciteSetBstMidEndSepPunct[3]{}
\providecommand*\mciteSetBstSublistLabelBeginEnd[3]{}
\providecommand*\EndOfBibitem{}
\mciteSetBstSublistMode{f}
\mciteSetBstMaxWidthForm{subitem}{(\alph{mcitesubitemcount})}
\mciteSetBstSublistLabelBeginEnd
  {\mcitemaxwidthsubitemform\space}
  {\relax}
  {\relax}

\bibitem[Hamann \latin{et~al.}(1979)Hamann, Schl\"{u}ter, and
  Chiang]{hamann1979}
Hamann,~D.~R.; Schl\"{u}ter,~M.; Chiang,~C. Norm-Conserving Pseudopotentials.
  \emph{Phys. Rev. Lett.} \textbf{1979}, \emph{43}, 1494--1497\relax
\mciteBstWouldAddEndPuncttrue
\mciteSetBstMidEndSepPunct{\mcitedefaultmidpunct}
{\mcitedefaultendpunct}{\mcitedefaultseppunct}\relax
\EndOfBibitem
\bibitem[Troullier and Martins(1991)Troullier, and Martins]{troullier1991}
Troullier,~N.; Martins,~J.~L. Efficient pseudopotentials for plane-wave
  calculations. \emph{Phys. Rev. B} \textbf{1991}, \emph{43}, 1993--2006\relax
\mciteBstWouldAddEndPuncttrue
\mciteSetBstMidEndSepPunct{\mcitedefaultmidpunct}
{\mcitedefaultendpunct}{\mcitedefaultseppunct}\relax
\EndOfBibitem
\bibitem[Vanderbilt(1990)]{vanderbilt1990}
Vanderbilt,~D. Soft self-consistent pseudopotentials in a generalized
  eigenvalue formalism. \emph{Phys. Rev. B} \textbf{1990}, \emph{41},
  7892--7895\relax
\mciteBstWouldAddEndPuncttrue
\mciteSetBstMidEndSepPunct{\mcitedefaultmidpunct}
{\mcitedefaultendpunct}{\mcitedefaultseppunct}\relax
\EndOfBibitem
\bibitem[Bl{\"o}chl \latin{et~al.}(2005)Bl{\"o}chl, K{\"a}stner, and
  F{\"o}rst]{Blochl2005}
Bl{\"o}chl,~P.~E.; K{\"a}stner,~J.; F{\"o}rst,~C.~J. In \emph{Handbook of
  Materials Modeling: Methods}; Yip,~S., Ed.; Springer Netherlands: Dordrecht,
  2005; pp 93--119\relax
\mciteBstWouldAddEndPuncttrue
\mciteSetBstMidEndSepPunct{\mcitedefaultmidpunct}
{\mcitedefaultendpunct}{\mcitedefaultseppunct}\relax
\EndOfBibitem
\bibitem[Kresse and Joubert(1999)Kresse, and Joubert]{kresse1999}
Kresse,~G.; Joubert,~D. From ultrasoft pseudopotentials to the projector
  augmented-wave method. \emph{Phys. Rev. B} \textbf{1999}, \emph{59},
  1758--1775\relax
\mciteBstWouldAddEndPuncttrue
\mciteSetBstMidEndSepPunct{\mcitedefaultmidpunct}
{\mcitedefaultendpunct}{\mcitedefaultseppunct}\relax
\EndOfBibitem
\bibitem[Kresse and Hafner(1993)Kresse, and Hafner]{vasp:prb_1993}
Kresse,~G.; Hafner,~J. Ab initio molecular dynamics for liquid metals.
  \emph{Phys. Rev. B} \textbf{1993}, \emph{47}, 558--561\relax
\mciteBstWouldAddEndPuncttrue
\mciteSetBstMidEndSepPunct{\mcitedefaultmidpunct}
{\mcitedefaultendpunct}{\mcitedefaultseppunct}\relax
\EndOfBibitem
\bibitem[Kresse and Furthm\"{u}ller(1996)Kresse, and
  Furthm\"{u}ller]{vasp:1996}
Kresse,~G.; Furthm\"{u}ller,~J. Efficiency of ab-initio total energy
  calculations for metals and semiconductors using a plane-wave basis set.
  \emph{Comput. Mat. Sci.} \textbf{1996}, \emph{6}, 15--50\relax
\mciteBstWouldAddEndPuncttrue
\mciteSetBstMidEndSepPunct{\mcitedefaultmidpunct}
{\mcitedefaultendpunct}{\mcitedefaultseppunct}\relax
\EndOfBibitem
\bibitem[Wimmer \latin{et~al.}(1981)Wimmer, Krakauer, Weinert, and
  Freeman]{FLAPW}
Wimmer,~E.; Krakauer,~H.; Weinert,~M.; Freeman,~A.~J. Full-potential
  self-consistent linearized-augmented-plane-wave method for calculating the
  electronic structure of molecules and surfaces: ${\mathrm{O}}_{2}$ molecule.
  \emph{Phys. Rev. B} \textbf{1981}, \emph{24}, 864--875\relax
\mciteBstWouldAddEndPuncttrue
\mciteSetBstMidEndSepPunct{\mcitedefaultmidpunct}
{\mcitedefaultendpunct}{\mcitedefaultseppunct}\relax
\EndOfBibitem
\bibitem[Hill(2013)]{Hill_gaussian}
Hill,~J.~G. Gaussian basis sets for molecular applications. \emph{Int. J.
  Quantum Chem.} \textbf{2013}, \emph{113}, 21--34\relax
\mciteBstWouldAddEndPuncttrue
\mciteSetBstMidEndSepPunct{\mcitedefaultmidpunct}
{\mcitedefaultendpunct}{\mcitedefaultseppunct}\relax
\EndOfBibitem
\bibitem[Jensen(2013)]{jensen:atomic_basis_review_2013}
Jensen,~F. Atomic orbital basis sets. \emph{Wiley Interdiscip. Rev. Comput.
  Mol. Sci.} \textbf{2013}, \emph{3}, 273--295\relax
\mciteBstWouldAddEndPuncttrue
\mciteSetBstMidEndSepPunct{\mcitedefaultmidpunct}
{\mcitedefaultendpunct}{\mcitedefaultseppunct}\relax
\EndOfBibitem
\bibitem[Lejaeghere \latin{et~al.}(2016)Lejaeghere, Bihlmayer, Bj{\"o}rkman,
  Blaha, Bl{\"u}gel, Blum, Caliste, Castelli, Clark, Corso, de~Gironcoli,
  Deutsch, Dewhurst, Marco, Draxl, Du{\l}ak, Eriksson, Flores-Livas, Garrity,
  Genovese, Giannozzi, Giantomassi, Goedecker, Gonze, Gr\r{a}n\"{a}s, Gross,
  Gulans, Gygi, Hamann, Hasnip, Holzwarth, Iu\c{s}an, Jochym, Jollet, Jones,
  Kresse, Koepernik, K\"{u}\c{c}\"{u}kbenli, Kvashnin, Locht, Lubeck, Marsman,
  Marzari, Nitzsche, Nordström, Ozaki, Paulatto, Pickard, Poelmans, Probert,
  Refson, Richter, Rignanese, Saha, Scheffler, Schlipf, Schwarz, Sharma,
  Tavazza, Thunstr\"{o}m, Tkatchenko, Torrent, Vanderbilt, van Setten,
  Speybroeck, Wills, Yates, Zhang, and Cottenier]{lejaeghere2016}
Lejaeghere,~K.; Bihlmayer,~G.; Bj{\"o}rkman,~T.; Blaha,~P.; Bl{\"u}gel,~S.;
  Blum,~V.; Caliste,~D.; Castelli,~I.~E.; Clark,~S.~J.; Corso,~A.~D.;
  de~Gironcoli,~S.; Deutsch,~T.; Dewhurst,~J.~K.; Marco,~I.~D.; Draxl,~C.;
  Du{\l}ak,~M.; Eriksson,~O.; Flores-Livas,~J.~A.; Garrity,~K.~F.;
  Genovese,~L.; Giannozzi,~P.; Giantomassi,~M.; Goedecker,~S.; Gonze,~X.;
  Gr\r{a}n\"{a}s,~O.; Gross,~E. K.~U.; Gulans,~A.; Gygi,~F.; Hamann,~D.~R.;
  Hasnip,~P.~J.; Holzwarth,~N. A.~W.; Iu\c{s}an,~D.; Jochym,~D.~B.; Jollet,~F.;
  Jones,~D.; Kresse,~G.; Koepernik,~K.; K\"{u}\c{c}\"{u}kbenli,~E.;
  Kvashnin,~Y.~O.; Locht,~I. L.~M.; Lubeck,~S.; Marsman,~M.; Marzari,~N.;
  Nitzsche,~U.; Nordström,~L.; Ozaki,~T.; Paulatto,~L.; Pickard,~C.~J.;
  Poelmans,~W.; Probert,~M. I.~J.; Refson,~K.; Richter,~M.; Rignanese,~G.-M.;
  Saha,~S.; Scheffler,~M.; Schlipf,~M.; Schwarz,~K.; Sharma,~S.; Tavazza,~F.;
  Thunstr\"{o}m,~P.; Tkatchenko,~A.; Torrent,~M.; Vanderbilt,~D.; van
  Setten,~M.~J.; Speybroeck,~V.~V.; Wills,~J.~M.; Yates,~J.~R.; Zhang,~G.-X.;
  Cottenier,~S. Reproducibility in density functional theory calculations of
  solids. \emph{Science} \textbf{2016}, \emph{351}, aad3000\relax
\mciteBstWouldAddEndPuncttrue
\mciteSetBstMidEndSepPunct{\mcitedefaultmidpunct}
{\mcitedefaultendpunct}{\mcitedefaultseppunct}\relax
\EndOfBibitem
\bibitem[Lejaeghere \latin{et~al.}(2014)Lejaeghere, Speybroeck, Oost, and
  Cottenier]{lejaeghere2014}
Lejaeghere,~K.; Speybroeck,~V.~V.; Oost,~G.~V.; Cottenier,~S. Error Estimates
  for Solid-State Density-Functional Theory Predictions: An Overview by Means
  of the Ground-State Elemental Crystals. \emph{Crit. Rev. Solid State Mater.
  Sci.} \textbf{2014}, \emph{39}, 1--24\relax
\mciteBstWouldAddEndPuncttrue
\mciteSetBstMidEndSepPunct{\mcitedefaultmidpunct}
{\mcitedefaultendpunct}{\mcitedefaultseppunct}\relax
\EndOfBibitem
\bibitem[Paier \latin{et~al.}(2005)Paier, Hirschl, Marsman, and
  Kresse]{Kresse:PBE_PBE0_G2_1_test_jcp_2005}
Paier,~J.; Hirschl,~R.; Marsman,~M.; Kresse,~G. The Perdew–Burke–Ernzerhof
  exchange-correlation functional applied to the G2-1 test set using a
  plane-wave basis set. \emph{J. Chem. Phys.} \textbf{2005}, \emph{122},
  234102\relax
\mciteBstWouldAddEndPuncttrue
\mciteSetBstMidEndSepPunct{\mcitedefaultmidpunct}
{\mcitedefaultendpunct}{\mcitedefaultseppunct}\relax
\EndOfBibitem
\bibitem[Pople \latin{et~al.}(1989)Pople, Head-Gordon, Fox, Raghavachari, and
  Curtiss]{pople1989}
Pople,~J.~A.; Head-Gordon,~M.; Fox,~D.~J.; Raghavachari,~K.; Curtiss,~L.~A.
  {Gaussian-1 theory: A general procedure for prediction of molecular
  energies}. \emph{J. Chem. Phys.} \textbf{1989}, \emph{90}, 5622--5629\relax
\mciteBstWouldAddEndPuncttrue
\mciteSetBstMidEndSepPunct{\mcitedefaultmidpunct}
{\mcitedefaultendpunct}{\mcitedefaultseppunct}\relax
\EndOfBibitem
\bibitem[Curtiss \latin{et~al.}(1990)Curtiss, Jones, Trucks, Raghavachari, and
  Pople]{curtiss1990}
Curtiss,~L.~A.; Jones,~C.; Trucks,~G.~W.; Raghavachari,~K.; Pople,~J.~A.
  {Gaussian-1 theory of molecular energies for second row compounds}. \emph{J.
  Chem. Phys.} \textbf{1990}, \emph{93}, 2537--2545\relax
\mciteBstWouldAddEndPuncttrue
\mciteSetBstMidEndSepPunct{\mcitedefaultmidpunct}
{\mcitedefaultendpunct}{\mcitedefaultseppunct}\relax
\EndOfBibitem
\bibitem[Perdew \latin{et~al.}(1996)Perdew, Burke, and
  Ernzerhof]{perdew:pbe_prl}
Perdew,~J.~P.; Burke,~K.; Ernzerhof,~M. Generalized Gradient Approximation Made
  Simple. \emph{Phys. Rev. Lett.} \textbf{1996}, \emph{77}, 3865--3868\relax
\mciteBstWouldAddEndPuncttrue
\mciteSetBstMidEndSepPunct{\mcitedefaultmidpunct}
{\mcitedefaultendpunct}{\mcitedefaultseppunct}\relax
\EndOfBibitem
\bibitem[Perdew \latin{et~al.}(1996)Perdew, Ernzerhof, and Burke]{pbe0}
Perdew,~J.~P.; Ernzerhof,~M.; Burke,~K. Rationale for mixing exact exchange
  with density functional approximations. \emph{J. Chem. Phys.} \textbf{1996},
  \emph{105}, 9982--9985\relax
\mciteBstWouldAddEndPuncttrue
\mciteSetBstMidEndSepPunct{\mcitedefaultmidpunct}
{\mcitedefaultendpunct}{\mcitedefaultseppunct}\relax
\EndOfBibitem
\bibitem[Woon and Dunning(1994)Woon, and Dunning]{dunning:basis_set}
Woon,~D.~E.; Dunning,~T.~H. Gaussian basis sets for use in correlated molecular
  calculations. IV. Calculation of static electrical response properties.
  \emph{J. Chem. Phys.} \textbf{1994}, \emph{100}, 2975--2988\relax
\mciteBstWouldAddEndPuncttrue
\mciteSetBstMidEndSepPunct{\mcitedefaultmidpunct}
{\mcitedefaultendpunct}{\mcitedefaultseppunct}\relax
\EndOfBibitem
\bibitem[Dunning \latin{et~al.}(2001)Dunning, Peterson, and
  Wilson]{dunning:d_func_basis_set}
Dunning,~T.~H.; Peterson,~K.~A.; Wilson,~A.~K. Gaussian basis sets for use in
  correlated molecular calculations. X. The atoms aluminum through argon
  revisited. \emph{J. Chem. Phys.} \textbf{2001}, \emph{114}, 9244--9253\relax
\mciteBstWouldAddEndPuncttrue
\mciteSetBstMidEndSepPunct{\mcitedefaultmidpunct}
{\mcitedefaultendpunct}{\mcitedefaultseppunct}\relax
\EndOfBibitem
\bibitem[Maggio \latin{et~al.}(2017)Maggio, Liu, van Setten, and
  Kresse]{kresse:gw_jctc_2017}
Maggio,~E.; Liu,~P.; van Setten,~M.~J.; Kresse,~G. GW100: A Plane Wave
  Perspective for Small Molecules. \emph{J. Chem. Theory Comput.}
  \textbf{2017}, \emph{13}, 635--648\relax
\mciteBstWouldAddEndPuncttrue
\mciteSetBstMidEndSepPunct{\mcitedefaultmidpunct}
{\mcitedefaultendpunct}{\mcitedefaultseppunct}\relax
\EndOfBibitem
\bibitem[Hedin(1965)]{Hedin:gw_PhysRev}
Hedin,~L. New Method for Calculating the One-Particle Green's Function with
  Application to the Electron-Gas Problem. \emph{Phys. Rev.} \textbf{1965},
  \emph{139}, A796--A823\relax
\mciteBstWouldAddEndPuncttrue
\mciteSetBstMidEndSepPunct{\mcitedefaultmidpunct}
{\mcitedefaultendpunct}{\mcitedefaultseppunct}\relax
\EndOfBibitem
\bibitem[van Setten \latin{et~al.}(2015)van Setten, Caruso, Sharifzadeh, Ren,
  Scheffler, Liu, Lischner, Lin, Deslippe, Louie, Yang, Weigend, Neaton, Evers,
  and Rinke]{setten2015}
van Setten,~M.~J.; Caruso,~F.; Sharifzadeh,~S.; Ren,~X.; Scheffler,~M.;
  Liu,~F.; Lischner,~J.; Lin,~L.; Deslippe,~J.~R.; Louie,~S.~G.; Yang,~C.;
  Weigend,~F.; Neaton,~J.~B.; Evers,~F.; Rinke,~P. GW100: Benchmarking G0W0 for
  Molecular Systems. \emph{J. Chem. Theory Comput.} \textbf{2015}, \emph{11},
  5665--5687\relax
\mciteBstWouldAddEndPuncttrue
\mciteSetBstMidEndSepPunct{\mcitedefaultmidpunct}
{\mcitedefaultendpunct}{\mcitedefaultseppunct}\relax
\EndOfBibitem
\bibitem[Adllan and Corso(2011)Adllan, and Corso]{adllan2011}
Adllan,~A.~A.; Corso,~A.~D. Ultrasoft pseudopotentials and projector
  augmented-wave data sets: application to diatomic molecules. \emph{J.
  Condens. Matter Phys.} \textbf{2011}, \emph{23}, 425501\relax
\mciteBstWouldAddEndPuncttrue
\mciteSetBstMidEndSepPunct{\mcitedefaultmidpunct}
{\mcitedefaultendpunct}{\mcitedefaultseppunct}\relax
\EndOfBibitem
\bibitem[Tkatchenko and Scheffler(2009)Tkatchenko, and
  Scheffler]{Tkatchenk:TS-HI}
Tkatchenko,~A.; Scheffler,~M. Accurate Molecular Van Der Waals Interactions
  from Ground-State Electron Density and Free-Atom Reference Data. \emph{Phys.
  Rev. Lett.} \textbf{2009}, \emph{102}, 073005\relax
\mciteBstWouldAddEndPuncttrue
\mciteSetBstMidEndSepPunct{\mcitedefaultmidpunct}
{\mcitedefaultendpunct}{\mcitedefaultseppunct}\relax
\EndOfBibitem
\bibitem[Hofierka and Klimes()Hofierka, and Klimes]{jiri2019pq_poster}
Hofierka,~J.; Klimes,~J. Understanding precision of binding energies of
  molecules and molecular solids.
  \url{http://quantum.karlov.mff.cuni.cz/~jklimes/pq0519.pdf}\relax
\mciteBstWouldAddEndPuncttrue
\mciteSetBstMidEndSepPunct{\mcitedefaultmidpunct}
{\mcitedefaultendpunct}{\mcitedefaultseppunct}\relax
\EndOfBibitem
\bibitem[Witte \latin{et~al.}(2019)Witte, Neaton, and Head-Gordon]{witte2019}
Witte,~J.; Neaton,~J.~B.; Head-Gordon,~M. Push it to the limit: comparing
  periodic and local approaches to density functional theory for intermolecular
  interactions. \emph{Mol. Phys.} \textbf{2019}, \emph{117}, 1298--1305\relax
\mciteBstWouldAddEndPuncttrue
\mciteSetBstMidEndSepPunct{\mcitedefaultmidpunct}
{\mcitedefaultendpunct}{\mcitedefaultseppunct}\relax
\EndOfBibitem
\bibitem[Perdew and Wang(1992)Perdew, and Wang]{spw92_functional}
Perdew,~J.~P.; Wang,~Y. Accurate and simple analytic representation of the
  electron-gas correlation energy. \emph{Phys. Rev. B} \textbf{1992},
  \emph{45}, 13244--13249\relax
\mciteBstWouldAddEndPuncttrue
\mciteSetBstMidEndSepPunct{\mcitedefaultmidpunct}
{\mcitedefaultendpunct}{\mcitedefaultseppunct}\relax
\EndOfBibitem
\bibitem[Jure\v{c}ka \latin{et~al.}(2006)Jure\v{c}ka, \v{S}poner,
  \v{C}ern\'{y}, and Hobza]{jurecka2006}
Jure\v{c}ka,~P.; \v{S}poner,~J.; \v{C}ern\'{y},~J.; Hobza,~P. Benchmark
  database of accurate (MP2 and CCSD(T) complete basis set limit) interaction
  energies of small model complexes{,} DNA base pairs{,} and amino acid pairs.
  \emph{Phys. Chem. Chem. Phys.} \textbf{2006}, \emph{8}, 1985--1993\relax
\mciteBstWouldAddEndPuncttrue
\mciteSetBstMidEndSepPunct{\mcitedefaultmidpunct}
{\mcitedefaultendpunct}{\mcitedefaultseppunct}\relax
\EndOfBibitem
\bibitem[Tosoni \latin{et~al.}(2007)Tosoni, Tuma, Sauer, Civalleri, and
  Ugliengo]{Civalleri:jcp_2007}
Tosoni,~S.; Tuma,~C.; Sauer,~J.; Civalleri,~B.; Ugliengo,~P. {A comparison
  between plane wave and Gaussian-type orbital basis sets for hydrogen bonded
  systems: Formic acid as a test case}. \emph{J. Chem. Phys.} \textbf{2007},
  \emph{127}, 154102\relax
\mciteBstWouldAddEndPuncttrue
\mciteSetBstMidEndSepPunct{\mcitedefaultmidpunct}
{\mcitedefaultendpunct}{\mcitedefaultseppunct}\relax
\EndOfBibitem
\bibitem[\v{R}ez\'{a}\v{c} \latin{et~al.}(2011)\v{R}ez\'{a}\v{c}, Riley, and
  Hobza]{S66_database}
\v{R}ez\'{a}\v{c},~J.; Riley,~K.~E.; Hobza,~P. S66: A Well-balanced Database of
  Benchmark Interaction Energies Relevant to Biomolecular Structures. \emph{J.
  Chem. Theory Comput.} \textbf{2011}, \emph{7}, 2427--2438\relax
\mciteBstWouldAddEndPuncttrue
\mciteSetBstMidEndSepPunct{\mcitedefaultmidpunct}
{\mcitedefaultendpunct}{\mcitedefaultseppunct}\relax
\EndOfBibitem
\bibitem[Kresse and Furthm\"uller(1996)Kresse, and Furthm\"uller]{kresse1996}
Kresse,~G.; Furthm\"uller,~J. Efficient iterative schemes for ab initio
  total-energy calculations using a plane-wave basis set. \emph{Phys. Rev. B}
  \textbf{1996}, \emph{54}, 11169--11186\relax
\mciteBstWouldAddEndPuncttrue
\mciteSetBstMidEndSepPunct{\mcitedefaultmidpunct}
{\mcitedefaultendpunct}{\mcitedefaultseppunct}\relax
\EndOfBibitem
\bibitem[\v{R}ez\'{a}\v{c} \latin{et~al.}(2008)\v{R}ez\'{a}\v{c}, Jure\v{c}ka,
  Riley, \v{C}ern\'{y}, Valdesa, Pluh\'{a}\v{c}kov\'{a}, Karel~Berka,
  Pito\v{n}\'{a}k, Vondr\'{a}\v{s}ek, and Hobza]{begdb_website}
\v{R}ez\'{a}\v{c},~J.; Jure\v{c}ka,~P.; Riley,~K.~E.; \v{C}ern\'{y},~J.;
  Valdesa,~H.; Pluh\'{a}\v{c}kov\'{a},~K.; Karel~Berka,~T.~v.;
  Pito\v{n}\'{a}k,~M.; Vondr\'{a}\v{s}ek,~J.; Hobza,~P. Quantum Chemical
  Benchmark Energy and Geometry Database for Molecular Clusters and Complex
  Molecular Systems (www.begdb.com): A Users Manual and Examples.
  \emph{Collect. Czech. Chem. Commun.} \textbf{2008}, \emph{73},
  1261--1270\relax
\mciteBstWouldAddEndPuncttrue
\mciteSetBstMidEndSepPunct{\mcitedefaultmidpunct}
{\mcitedefaultendpunct}{\mcitedefaultseppunct}\relax
\EndOfBibitem
\bibitem[Makov and Payne(1995)Makov, and
  Payne]{Makov_Payne:PBC_dipole_correction}
Makov,~G.; Payne,~M.~C. Periodic boundary conditions in ab initio calculations.
  \emph{Phys. Rev. B} \textbf{1995}, \emph{51}, 4014--4022\relax
\mciteBstWouldAddEndPuncttrue
\mciteSetBstMidEndSepPunct{\mcitedefaultmidpunct}
{\mcitedefaultendpunct}{\mcitedefaultseppunct}\relax
\EndOfBibitem
\bibitem[Gygi and Baldereschi(1986)Gygi, and Baldereschi]{gygi1986}
Gygi,~F.; Baldereschi,~A. Self-consistent Hartree-Fock and screened-exchange
  calculations in solids: Application to silicon. \emph{Phys. Rev. B}
  \textbf{1986}, \emph{34}, 4405--4408\relax
\mciteBstWouldAddEndPuncttrue
\mciteSetBstMidEndSepPunct{\mcitedefaultmidpunct}
{\mcitedefaultendpunct}{\mcitedefaultseppunct}\relax
\EndOfBibitem
\bibitem[Hofierka and Klime{\v{s}}(2021)Hofierka, and
  Klime{\v{s}}]{jiri:elec_stru_pbc_mbe}
Hofierka,~J.; Klime{\v{s}},~J. Binding energies of molecular solids from
  fragment and periodic approaches. \emph{Electron. Struct.} \textbf{2021},
  \emph{3}, 034010\relax
\mciteBstWouldAddEndPuncttrue
\mciteSetBstMidEndSepPunct{\mcitedefaultmidpunct}
{\mcitedefaultendpunct}{\mcitedefaultseppunct}\relax
\EndOfBibitem
\bibitem[Bader(1990)]{Bader:book}
Bader,~R. F.~W. \emph{Atoms in Molecules -- A Quantum Theory}; Oxford
  University Press: Oxford, 1990\relax
\mciteBstWouldAddEndPuncttrue
\mciteSetBstMidEndSepPunct{\mcitedefaultmidpunct}
{\mcitedefaultendpunct}{\mcitedefaultseppunct}\relax
\EndOfBibitem
\bibitem[Bultinck \latin{et~al.}(2007)Bultinck, Van~Alsenoy, Ayers, and
  Carb{\'o}-Dorca]{bultinck:hirshfeld_algorithm}
Bultinck,~P.; Van~Alsenoy,~C.; Ayers,~P.~W.; Carb{\'o}-Dorca,~R. Critical
  analysis and extension of the Hirshfeld atoms in molecules. \emph{J. Chem.
  Phys.} \textbf{2007}, \emph{126}, 144111\relax
\mciteBstWouldAddEndPuncttrue
\mciteSetBstMidEndSepPunct{\mcitedefaultmidpunct}
{\mcitedefaultendpunct}{\mcitedefaultseppunct}\relax
\EndOfBibitem
\bibitem[Bultinck \latin{et~al.}(2007)Bultinck, Ayers, Fias, Tiels, and {Van
  Alsenoy}]{BULTINCK:cpl_2007}
Bultinck,~P.; Ayers,~P.~W.; Fias,~S.; Tiels,~K.; {Van Alsenoy},~C. Uniqueness
  and basis set dependence of iterative Hirshfeld charges. \emph{Chem. Phys.
  Lett.} \textbf{2007}, \emph{444}, 205--208\relax
\mciteBstWouldAddEndPuncttrue
\mciteSetBstMidEndSepPunct{\mcitedefaultmidpunct}
{\mcitedefaultendpunct}{\mcitedefaultseppunct}\relax
\EndOfBibitem
\bibitem[Bu\v{c}ko \latin{et~al.}(2013)Bu\v{c}ko, Leb\'{e}gue, Hafner, and
  \'{A}ngy\'{a}n]{Angyan:ts_hi_explanation_jctc_2013}
Bu\v{c}ko,~T.; Leb\'{e}gue,~S.; Hafner,~J.; \'{A}ngy\'{a}n,~J.~G. Improved
  Density Dependent Correction for the Description of London Dispersion Forces.
  \emph{J. Chem. Theory Comput.} \textbf{2013}, \emph{9}, 4293--4299\relax
\mciteBstWouldAddEndPuncttrue
\mciteSetBstMidEndSepPunct{\mcitedefaultmidpunct}
{\mcitedefaultendpunct}{\mcitedefaultseppunct}\relax
\EndOfBibitem
\bibitem[Henkelman \latin{et~al.}(2006)Henkelman, Arnaldsson, and
  J{'o}nsson]{bader_code:Henkelman_2006}
Henkelman,~G.; Arnaldsson,~A.; J{'o}nsson,~H. A fast and robust algorithm for
  Bader decomposition of charge density. \emph{Comput. Mater. Sci.}
  \textbf{2006}, \emph{36}, 354--360\relax
\mciteBstWouldAddEndPuncttrue
\mciteSetBstMidEndSepPunct{\mcitedefaultmidpunct}
{\mcitedefaultendpunct}{\mcitedefaultseppunct}\relax
\EndOfBibitem
\bibitem[Sanville \latin{et~al.}(2007)Sanville, Kenny, Smith, and
  Henkelman]{bader_code:henkelman_2007}
Sanville,~E.; Kenny,~S.~D.; Smith,~R.; Henkelman,~G. Improved grid-based
  algorithm for Bader charge allocation. \emph{J. Comput. Chem.} \textbf{2007},
  \emph{28}, 899--908\relax
\mciteBstWouldAddEndPuncttrue
\mciteSetBstMidEndSepPunct{\mcitedefaultmidpunct}
{\mcitedefaultendpunct}{\mcitedefaultseppunct}\relax
\EndOfBibitem
\bibitem[Tang \latin{et~al.}(2009)Tang, Sanville, and
  Henkelman]{bader_code:henkelman_2009}
Tang,~W.; Sanville,~E.; Henkelman,~G. A grid-based Bader analysis algorithm
  without lattice bias. \emph{J. Phys. Condens. Matter} \textbf{2009},
  \emph{21}, 084204\relax
\mciteBstWouldAddEndPuncttrue
\mciteSetBstMidEndSepPunct{\mcitedefaultmidpunct}
{\mcitedefaultendpunct}{\mcitedefaultseppunct}\relax
\EndOfBibitem
\bibitem[Yu and Trinkle(2011)Yu, and Trinkle]{Bader_code:Trinkle_2011}
Yu,~M.; Trinkle,~D.~R. Accurate and efficient algorithm for Bader charge
  integration. \emph{J. Chem. Phys.} \textbf{2011}, \emph{134}, 064111\relax
\mciteBstWouldAddEndPuncttrue
\mciteSetBstMidEndSepPunct{\mcitedefaultmidpunct}
{\mcitedefaultendpunct}{\mcitedefaultseppunct}\relax
\EndOfBibitem
\bibitem[Balasubramani \latin{et~al.}(2020)Balasubramani, Chen, Coriani,
  Diedenhofen, Frank, Franzke, Furche, Grotjahn, Harding, H\"{a}ttig, Hellweg,
  Helmich-Paris, Holzer, Huniar, Kaupp, Marefat~Khah, Karbalaei~Khani,
  M\"{u}ller, Mack, Nguyen, Parker, Perlt, Rappoport, Reiter, Roy, R\"{u}ckert,
  Schmitz, Sierka, Tapavicza, Tew, van W\"{u}llen, Voora, Weigend,
  Wody\'{n}ski, and Yu]{TURBOMOLE}
Balasubramani,~S.~G.; Chen,~G.~P.; Coriani,~S.; Diedenhofen,~M.; Frank,~M.~S.;
  Franzke,~Y.~J.; Furche,~F.; Grotjahn,~R.; Harding,~M.~E.; H\"{a}ttig,~C.;
  Hellweg,~A.; Helmich-Paris,~B.; Holzer,~C.; Huniar,~U.; Kaupp,~M.;
  Marefat~Khah,~A.; Karbalaei~Khani,~S.; M\"{u}ller,~T.; Mack,~F.;
  Nguyen,~B.~D.; Parker,~S.~M.; Perlt,~E.; Rappoport,~D.; Reiter,~K.; Roy,~S.;
  R\"{u}ckert,~M.; Schmitz,~G.; Sierka,~M.; Tapavicza,~E.; Tew,~D.~P.; van
  W\"{u}llen,~C.; Voora,~V.~K.; Weigend,~F.; Wody\'{n}ski,~A.; Yu,~J.~M.
  {TURBOMOLE: Modular program suite for ab initio quantum-chemical and
  condensed-matter simulations}. \emph{The Journal of Chemical Physics}
  \textbf{2020}, \emph{152}, 184107\relax
\mciteBstWouldAddEndPuncttrue
\mciteSetBstMidEndSepPunct{\mcitedefaultmidpunct}
{\mcitedefaultendpunct}{\mcitedefaultseppunct}\relax
\EndOfBibitem
\bibitem[{\L}azarski \latin{et~al.}(2015){\L}azarski, Burow, and
  Sierka]{density_fitting}
{\L}azarski,~R.; Burow,~A.~M.; Sierka,~M. Density Functional Theory for
  Molecular and Periodic Systems Using Density Fitting and Continuous Fast
  Multipole Methods. \emph{J. Chem. Theory Comput.} \textbf{2015}, \emph{11},
  3029--3041\relax
\mciteBstWouldAddEndPuncttrue
\mciteSetBstMidEndSepPunct{\mcitedefaultmidpunct}
{\mcitedefaultendpunct}{\mcitedefaultseppunct}\relax
\EndOfBibitem
\bibitem[Boys and Bernardi(1970)Boys, and Bernardi]{CP:boys_bernardi}
Boys,~S.; Bernardi,~F. The calculation of small molecular interactions by the
  differences of separate total energies. Some procedures with reduced errors.
  \emph{Mol. Phys.} \textbf{1970}, \emph{19}, 553--566\relax
\mciteBstWouldAddEndPuncttrue
\mciteSetBstMidEndSepPunct{\mcitedefaultmidpunct}
{\mcitedefaultendpunct}{\mcitedefaultseppunct}\relax
\EndOfBibitem
\bibitem[Jensen(2001)]{Jensen:pc_basis_1}
Jensen,~F. Polarization consistent basis sets: Principles. \emph{J. Chem.
  Phys.} \textbf{2001}, \emph{115}, 9113--9125\relax
\mciteBstWouldAddEndPuncttrue
\mciteSetBstMidEndSepPunct{\mcitedefaultmidpunct}
{\mcitedefaultendpunct}{\mcitedefaultseppunct}\relax
\EndOfBibitem
\bibitem[Jensen(2002)]{Jensen:pc_basis_2}
Jensen,~F. Polarization consistent basis sets. II. Estimating the Kohn–Sham
  basis set limit. \emph{J. Chem. Phys.} \textbf{2002}, \emph{116},
  7372--7379\relax
\mciteBstWouldAddEndPuncttrue
\mciteSetBstMidEndSepPunct{\mcitedefaultmidpunct}
{\mcitedefaultendpunct}{\mcitedefaultseppunct}\relax
\EndOfBibitem
\bibitem[Jeziorski \latin{et~al.}(1994)Jeziorski, Moszynski, and
  Szalewicz]{sapt:chem_rev}
Jeziorski,~B.; Moszynski,~R.; Szalewicz,~K. Perturbation Theory Approach to
  Intermolecular Potential Energy Surfaces of van der Waals Complexes.
  \emph{Chem. Rev.} \textbf{1994}, \emph{94}, 1887--1930\relax
\mciteBstWouldAddEndPuncttrue
\mciteSetBstMidEndSepPunct{\mcitedefaultmidpunct}
{\mcitedefaultendpunct}{\mcitedefaultseppunct}\relax
\EndOfBibitem
\bibitem[He{\ss}elmann(2018)]{hesselmann:sapt_exx_jctc_2018}
He{\ss}elmann,~A. DFT-SAPT Intermolecular Interaction Energies Employing
  Exact-Exchange Kohn–Sham Response Methods. \emph{J. Chem. Theory Comput.}
  \textbf{2018}, \emph{14}, 1943--1959\relax
\mciteBstWouldAddEndPuncttrue
\mciteSetBstMidEndSepPunct{\mcitedefaultmidpunct}
{\mcitedefaultendpunct}{\mcitedefaultseppunct}\relax
\EndOfBibitem
\bibitem[He{\ss}elmann \latin{et~al.}(2005)He{\ss}elmann, Jansen, and
  Sch\"{u}tz]{hesselmann:jcp_2005}
He{\ss}elmann,~A.; Jansen,~G.; Sch\"{u}tz,~M. Density-functional
  theory-symmetry-adapted intermolecular perturbation theory with density
  fitting: A new efficient method to study intermolecular interaction energies.
  \emph{J. Chem. Phys.} \textbf{2005}, \emph{122}, 014103\relax
\mciteBstWouldAddEndPuncttrue
\mciteSetBstMidEndSepPunct{\mcitedefaultmidpunct}
{\mcitedefaultendpunct}{\mcitedefaultseppunct}\relax
\EndOfBibitem
\bibitem[G\"{o}rling(1998)]{gorling:ijqc_exx_98}
G\"{o}rling,~A. Exact exchange kernel for time-dependent density-functional
  theory. \emph{Int. J. Quantum Chem.} \textbf{1998}, \emph{69}, 265--277\relax
\mciteBstWouldAddEndPuncttrue
\mciteSetBstMidEndSepPunct{\mcitedefaultmidpunct}
{\mcitedefaultendpunct}{\mcitedefaultseppunct}\relax
\EndOfBibitem
\bibitem[G\"orling(1998)]{gorling:exx_pra_98}
G\"orling,~A. Exact exchange-correlation kernel for dynamic response properties
  and excitation energies in density-functional theory. \emph{Phys. Rev. A}
  \textbf{1998}, \emph{57}, 3433--3436\relax
\mciteBstWouldAddEndPuncttrue
\mciteSetBstMidEndSepPunct{\mcitedefaultmidpunct}
{\mcitedefaultendpunct}{\mcitedefaultseppunct}\relax
\EndOfBibitem
\bibitem[Blanco \latin{et~al.}(2005)Blanco, Mart\'{i}n~Pend\'{a}s, and
  Francisco]{IQA_1}
Blanco,~M.~A.; Mart\'{i}n~Pend\'{a}s,~A.; Francisco,~E. Interacting Quantum
  Atoms: A Correlated Energy Decomposition Scheme Based on the Quantum Theory
  of Atoms in Molecules. \emph{J. Chem. Theory Comput.} \textbf{2005},
  \emph{1}, 1096--1109\relax
\mciteBstWouldAddEndPuncttrue
\mciteSetBstMidEndSepPunct{\mcitedefaultmidpunct}
{\mcitedefaultendpunct}{\mcitedefaultseppunct}\relax
\EndOfBibitem
\bibitem[Francisco \latin{et~al.}(2006)Francisco, Mart\'{i}n~Pend\'{a}s, and
  Blanco]{IQA_2}
Francisco,~E.; Mart\'{i}n~Pend\'{a}s,~A.; Blanco,~M.~A. A Molecular Energy
  Decomposition Scheme for Atoms in Molecules. \emph{J. Chem. Theory Comput.}
  \textbf{2006}, \emph{2}, 90--102\relax
\mciteBstWouldAddEndPuncttrue
\mciteSetBstMidEndSepPunct{\mcitedefaultmidpunct}
{\mcitedefaultendpunct}{\mcitedefaultseppunct}\relax
\EndOfBibitem
\bibitem[Werner \latin{et~al.}(2012)Werner, Knowles, Knizia, Manby, and
  Sch\"{u}tz]{molpro_1}
Werner,~H.-J.; Knowles,~P.~J.; Knizia,~G.; Manby,~F.~R.; Sch\"{u}tz,~M. Molpro:
  a general-purpose quantum chemistry program package. \emph{Wiley Interdiscip.
  Rev. Comput. Mol. Sci.} \textbf{2012}, \emph{2}, 242--253\relax
\mciteBstWouldAddEndPuncttrue
\mciteSetBstMidEndSepPunct{\mcitedefaultmidpunct}
{\mcitedefaultendpunct}{\mcitedefaultseppunct}\relax
\EndOfBibitem
\bibitem[Werner \latin{et~al.}(2020)Werner, Knowles, Manby, Black, Doll,
  He{\ss}elmann, Kats, K\"{o}hn, Korona, Kreplin, Ma, Miller, Mitrushchenkov,
  Peterson, Polyak, Rauhut, and Sibaev]{molpro_2}
Werner,~H.-J.; Knowles,~P.~J.; Manby,~F.~R.; Black,~J.~A.; Doll,~K.;
  He{\ss}elmann,~A.; Kats,~D.; K\"{o}hn,~A.; Korona,~T.; Kreplin,~D.~A.;
  Ma,~Q.; Miller,~T.~F.; Mitrushchenkov,~A.; Peterson,~K.~A.; Polyak,~I.;
  Rauhut,~G.; Sibaev,~M. The Molpro quantum chemistry package. \emph{J. Chem.
  Phys.} \textbf{2020}, \emph{152}, 144107\relax
\mciteBstWouldAddEndPuncttrue
\mciteSetBstMidEndSepPunct{\mcitedefaultmidpunct}
{\mcitedefaultendpunct}{\mcitedefaultseppunct}\relax
\EndOfBibitem
\bibitem[Keith(2016)]{AIMAll}
Keith,~T.~A. {AIMAll} (Version 19.10.12). 2016\relax
\mciteBstWouldAddEndPuncttrue
\mciteSetBstMidEndSepPunct{\mcitedefaultmidpunct}
{\mcitedefaultendpunct}{\mcitedefaultseppunct}\relax
\EndOfBibitem
\bibitem[Witte \latin{et~al.}(2016)Witte, Neaton, and Head-Gordon]{witte2016}
Witte,~J.; Neaton,~J.~B.; Head-Gordon,~M. Push it to the limit: Characterizing
  the convergence of common sequences of basis sets for intermolecular
  interactions as described by density functional theory. \emph{J. Chem. Phys.}
  \textbf{2016}, \emph{144}, 194306\relax
\mciteBstWouldAddEndPuncttrue
\mciteSetBstMidEndSepPunct{\mcitedefaultmidpunct}
{\mcitedefaultendpunct}{\mcitedefaultseppunct}\relax
\EndOfBibitem
\bibitem[Bu\v{c}ko \latin{et~al.}(2013)Bu\v{c}ko, Leb\`egue, Hafner, and
  \'Angy\'an]{Bucko2013}
Bu\v{c}ko,~T.; Leb\`egue,~S.; Hafner,~J.; \'Angy\'an,~J.~G.
  Tkatchenko-Scheffler van der Waals correction method with and without
  self-consistent screening applied to solids. \emph{Phys. Rev. B}
  \textbf{2013}, \emph{87}, 064110\relax
\mciteBstWouldAddEndPuncttrue
\mciteSetBstMidEndSepPunct{\mcitedefaultmidpunct}
{\mcitedefaultendpunct}{\mcitedefaultseppunct}\relax
\EndOfBibitem
\bibitem[Thürlemann \latin{et~al.}(2022)Thürlemann, Böselt, and
  Riniker]{thurlemann2022}
Thürlemann,~M.; Böselt,~L.; Riniker,~S. Learning Atomic Multipoles:
  Prediction of the Electrostatic Potential with Equivariant Graph Neural
  Networks. \emph{J. Chem. Theory Comput.} \textbf{2022}, \emph{18},
  1701--1710, PMID: 35112866\relax
\mciteBstWouldAddEndPuncttrue
\mciteSetBstMidEndSepPunct{\mcitedefaultmidpunct}
{\mcitedefaultendpunct}{\mcitedefaultseppunct}\relax
\EndOfBibitem
\end{mcitethebibliography}

\begin{tocentry}
\includegraphics[width=7.0cm]{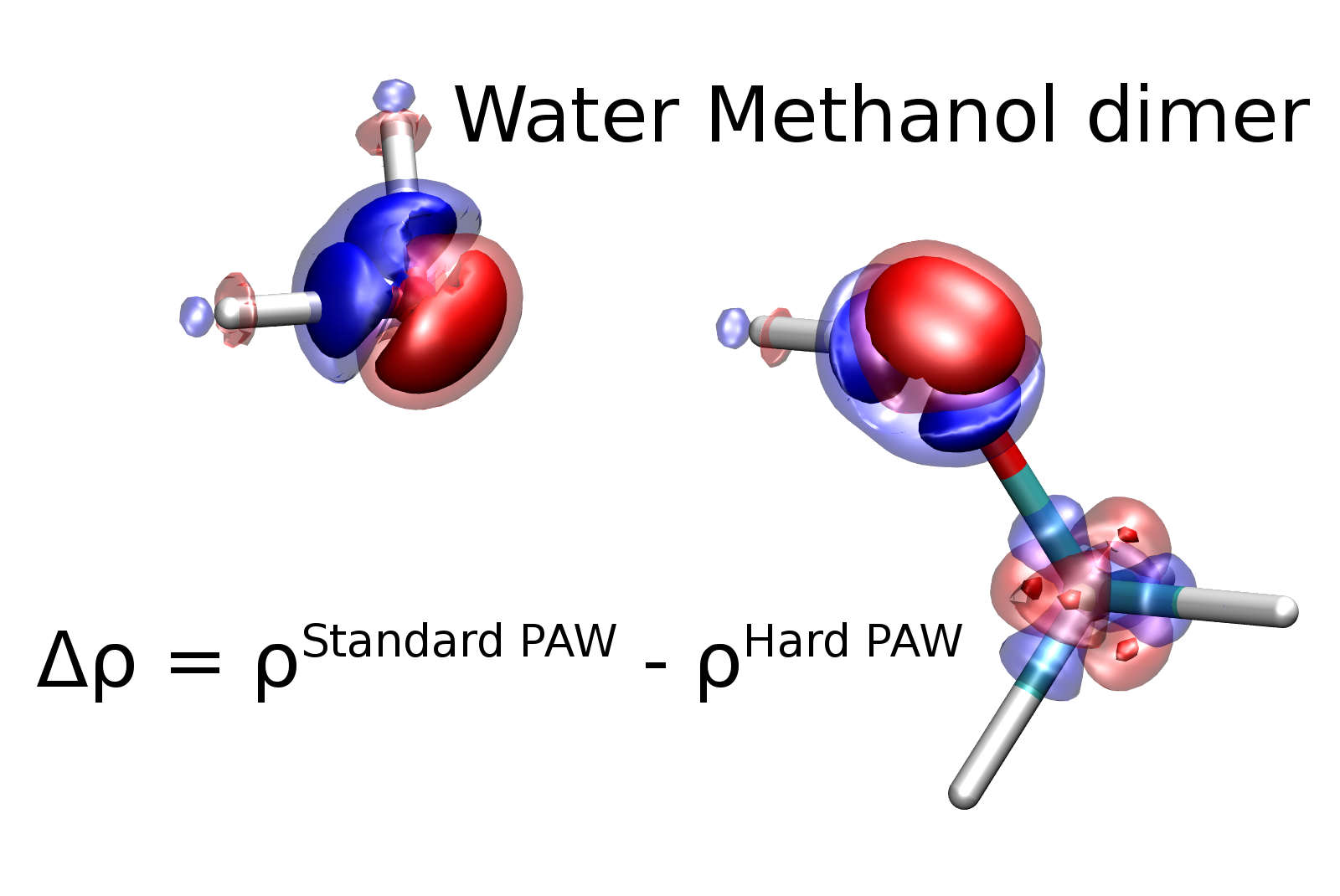}
\end{tocentry}

\end{document}